\documentclass[12pt,preprint]{aastex}
\usepackage{apjfonts}
\usepackage{graphicx}
\usepackage{natbib}

\newcommand{\kep}{$K_{ep}$}
\newcommand{\nh}{\bar{n}_{\rm H}}
\newcommand{\FGLnorth}{1FGL~J1801.3$-$2322c}
\newcommand{\FGLsouth}{1FGL~J1800.5$-$2359c}
\newcommand{\SourceN}{Source~N}
\newcommand{\SourceS}{Source~S}
\newcommand{\hii}{H~{\sc II}}

\slugcomment{To be re-submitted to ApJ: v5.7,  \today}

\shorttitle{\emph{Fermi} LAT Observations of SNR W28 }
\shortauthors{Abdo et al.}

\begin{document}


\title{\emph{Fermi} LAT Observations of the Supernova Remnant W28 (G6.4$-$0.1)}


%\author{The Fermi LAT Collaboration}
%\affil{Contact Author: Hideaki Katagiri, Hiroyasu Tajima, Takaaki Tanaka, Yasunobu Uchiyama
%Participants: Uehara, Nishino
%Affiliation list: Ryo Yamazaki, Yasuo Fukui, Hiroaki Yamamoto, Omar Tibolla}
%\email{katagiri@hep01.hepl.hiroshima-u.ac.jp}

%\title{Fermi LAT Observations of the Supernova Remnant W28 (G6.4−0.1)\\Author list created Thursday 11 Mar 2010 16:59 PST}

\author{
A.~A.~Abdo\altaffilmark{2,3}, 
M.~Ackermann\altaffilmark{4}, 
M.~Ajello\altaffilmark{4}, 
A.~Allafort\altaffilmark{4}, 
L.~Baldini\altaffilmark{5}, 
J.~Ballet\altaffilmark{6}, 
G.~Barbiellini\altaffilmark{7,8}, 
D.~Bastieri\altaffilmark{9,10}, 
K.~Bechtol\altaffilmark{4}, 
R.~Bellazzini\altaffilmark{5}
% ...
}
% ...
\altaffiltext{2}{Space Science Division, Naval Research Laboratory, Washington, DC 20375, USA}
\altaffiltext{3}{National Research Council Research Associate, National Academy of Sciences, Washington, DC 20001, USA}
\altaffiltext{4}{W. W. Hansen Experimental Physics Laboratory, Kavli Institute for Particle Astrophysics and Cosmology, Department of Physics and SLAC National Accelerator Laboratory, Stanford University, Stanford, CA 94305, USA}
\altaffiltext{5}{Istituto Nazionale di Fisica Nucleare, Sezione di Pisa, I-56127 Pisa, Italy}
\altaffiltext{6}{Laboratoire AIM, CEA-IRFU/CNRS/Universit\'e Paris Diderot, Service d'Astrophysique, CEA Saclay, 91191 Gif sur Yvette, France}
\altaffiltext{7}{Istituto Nazionale di Fisica Nucleare, Sezione di Trieste, I-34127 Trieste, Italy}
\altaffiltext{8}{Dipartimento di Fisica, Universit\`a di Trieste, I-34127 Trieste, Italy}
\altaffiltext{9}{Istituto Nazionale di Fisica Nucleare, Sezione di Padova, I-35131 Padova, Italy}
\altaffiltext{10}{Dipartimento di Fisica ``G. Galilei", Universit\`a di Padova, I-35131 Padova, Italy}
% ...
% This list is preliminary; the status is not yet "ready to submit"

\begin{abstract}
%Studies of cosmic-ray accelerations and interactions in supernova remnants (SNRs) are critical step to solve the mystery of the cosmic-ray origins.
%Observation of enhanced $\pi^0$ gamma rays in the GeV--TeV band from SNRs interacting with molecular clouds is one of the promising approaches to confirm hadronic nature of cosmic rays accelerated in SNRs.
We present detailed analysis of the two gamma-ray sources, \FGLnorth\ and \FGLsouth, 
 that have been found toward
the supernova remnant~(SNR) W28 with the Large Area Telescope~(LAT) on board the \emph{Fermi} Gamma-ray Space Telescope. 
 \FGLnorth\ is found to be an extended source within the boundary of SNR W28,
and to extensively overlap with the TeV gamma-ray 
source HESS~J1801$-$233, which is associated with a dense molecular cloud interacting with 
the supernova remnant. 
The gamma-ray spectrum measured with LAT from 0.2--100 GeV 
can be described by a broken power-law function with a break of
 $\sim$~1~GeV, and photon indices of 2.09~$\pm$~0.08~(stat)~$\pm$~0.28~(sys) below the break and 2.74~$\pm$~0.06~(stat)~$\pm$~0.09~(sys) above
 the break. 
Given the clear association between HESS~J1801$-$233 and the shocked
 molecular cloud and a smoothly connected spectrum in the GeV--TeV band, 
we consider the origin of the gamma-ray emission in both GeV and TeV ranges to be 
the interaction between particles accelerated in the SNR and the molecular cloud. 
The decay of neutral pions produced in interactions between accelerated hadrons 
and dense molecular gas provide a reasonable explanation for the broadband gamma-ray spectrum. 
% \FGLsouth\, located outside the southern boundary of SNR W28, is consistent with a point source. 
\FGLsouth\, located outside the southern boundary of SNR W28, 
 cannot be resolved.
An upper limit on the size of the gamma-ray emission was estimated to be
 $\sim$~16$'$
using events above $\sim$~2~GeV
under the assumption of a circular shape with uniform surface brightness.
It appears to coincide with the TeV source HESS~J1800$-$240B, 
which is considered to be associated with a dense molecular cloud that contains 
the ultra compact \hii\ region W28A2 (G5.89$-$0.39).
We found no significant gamma-ray emission in the LAT energy band at the positions of TeV sources HESS~J1800$-$230A and HESS~J1800$-$230C.
%The LAT data for HESS~J1800$-$230A combined with the TeV data points
 %indicate a spectral break around 100~GeV.
The LAT data for HESS~J1800$-$230A combined with the TeV data
 points indicate a spectral break between 10~GeV and 100~GeV.

\end{abstract}



\section{Introduction}
 Diffusive shock acceleration operating at supernova shock waves can  
distribute particles to very high energies 
with a power-law form having number index about 2 \citep[e.g.,][]{blandford87}.
It is generally expected that 
if a dense molecular cloud is overtaken by a supernova blast wave, the shocked molecular 
cloud can be illuminated by relativistic particles accelerated at supernova shocks 
\citep{Aharonian94}. If the accelerated particles are comprised mostly of protons, 
say $>100$ times more abundant than electrons like the 
observed Galactic cosmic rays, decays of neutral pions produced in inelastic collisions 
of the accelerated protons with dense gas are expected to be 
a dominant radiation component in the gamma-ray spectrum of the 
cosmic-ray-illuminated molecular cloud. 
Although an earlier attempt to detect TeV gamma-ray emission 
from supernova remnants (SNRs) that have evidence for molecular cloud interactions 
with the Whipple telescope failed \citep{Whipple98}, 
two archetypical SNRs interacting with molecular clouds, 
IC~443 \citep{Albert07, Acciari09} and W28 \citep{Aharonian08}, 
have been detected with the current generation of imaging atmospheric 
Cherenkov telescopes. However, the identities of the particles responsible
 for the TeV sources remains elusive. 

 The advent of the Large Area Telescope (LAT) onboard 
the \emph{Fermi} Gamma-ray Space Telescope has brought a new opportunity 
to study the gamma-ray emission from SNRs at GeV energies.
 LAT observations of middle-aged SNRs interacting with molecular clouds, W51C~\citep{LAT-W51C}, W44~\citep{LAT-W44}, and IC~443~\citep{LAT-IC443}, 
 have revealed bright extended 
gamma-ray sources coincident with the SNRs. 
The gamma-ray luminosity reaches $\sim 10^{36}\ \rm erg\ s^{-1}$, 
which immediately rules out an inverse-Compton origin of the GeV gamma rays 
since it requires total electron energy comparable to or larger than the typical kinetic energy released by a supernova explosion, $\sim 10^{51}$~erg.
The gamma-ray spectra of the three remnants exhibit remarkable 
spectral breaks at an energy of several GeV, making these SNRs much less luminous 
at TeV energies. This characteristic demonstrates the importance of observations 
at GeV energies. 
%In this paper, detailed analysis of LAT data in a field that contains 
%an archetypical SNR-cloud system, W28, will be presented.  

 W28 is a mixed-morphology SNR, characterized by 
center-filled thermal X-ray emission and shell-like radio morphology. 
In addition X-ray observations show
 limb-brightened shells in the northeast and southwest \citep{Rho02}.
The shell-like radio emission is prominent in the northeastern region
with slightly fainter emission at the northern boundaries \citep{Dubner00}.
Interactions of the SNR with molecular clouds~\citep{Wootten81} along
its northern and northeastern boundaries are traced by the high
concentration of 1720~MHz OH masers \citep{Frail94, Claussen97,Claussen99}, 
and  high density ($\nh > 10^3$~cm$^{-3}$) shocked gas \citep{Arikawa99, Reach05}.
The overall shape of W28 is elliptical with a size of $50'\times 45'$.
W28 is located within a complex, star-forming region along the Galactic plane toward the large \hii\ regions (M8 and M20) and young clusters~(e.g., NGC~6530)~\citep{goudis76}.
The observations of molecular lines place SNR W28 at a distance of $\sim 2$ kpc~\citep{Velazquez02}.
Estimates for its age vary between $3.5$ and $15\times 10^4$~yrs \citep{Kaspi93}.
W28 is considered to be an evolved remnant in the radiative
stage of SNR evolution \citep{Lozinskaya92}, which is supported by optical observations \citep{Lozinskaya74}. 
Measurements with Energetic Gamma-Ray Experiment Telescope (EGRET) 
onboard the \emph{Compton Gamma-ray Observatory} found a gamma-ray source, 3EG~J1800-2338~\citep{Hartman99} in the W28 field. 
However, its association to SNR W28 
remained unclear mainly due to large source location uncertainties from
EGRET.
A gamma-ray source is listed in the W28 field in the AGILE (Astro-rivelatore Gamma a Immagini LEggero) one year catalog~\citep{AGILEcatalog}. However, detailed analysis of this field is not published by AGILE yet.

 H.E.S.S. observations of the W28 field have revealed 
 four TeV gamma-ray sources positionally coincident with molecular clouds \citep{Aharonian08}: 
 HESS~J1801$-$233, located along the northeastern boundary of W28, and 
a complex of sources, HESS~J1800$-$240A, B and C,  located $\sim 30\arcmin$ south of 
SNR W28.
 HESS~J1801$-$233 coincides with a molecular cloud interacting with 
 SNR W28, providing one of the best examples of a cosmic-ray-illuminated cloud.
Understanding the origins of TeV emission in HESS~J1800$-$240ABC is  of particular 
interest; they may be due to cosmic rays that have diffused from W28.

In this paper, we report \emph{Fermi} LAT observations of gamma-ray sources in the W28 field in the GeV domain.
First, we give a brief description of the observation and gamma-ray selection in Section~\ref{sec:obs}.
The analysis procedures and results are explained in Section~\ref{sec:ana}, 
where the spatial extension and spectra of the LAT sources in the W28 field are described. 
Discussion is given in Section~\ref{sec:discuss}, 
followed by conclusions in Section~\ref{sec:conclusion}.

\section{OBSERVATION AND DATA REDUCTION}

\label{sec:obs}
%The LAT is the main instrument on \emph{Fermi}, sensitive to gamma rays in the energy band from $\sim$~20~MeV to $>$~300~GeV. 
The LAT is the main instrument on \emph{Fermi} sensitive to gamma rays.
The energy range spans from $\sim$~20~MeV to $>$~300~GeV, 
although as noted below in the present analysis we use only events with energies $>$~200~MeV.
It is an electron-positron pair production telescope, built with tungsten foils and silicon microstrip detectors to measure the arrival directions of incoming gamma rays, and a hodoscopic cesium iodide calorimeter to determine the gamma-ray energies.
They are surrounded by 89 segmented  plastic scintillators that serve as an anticoincidence detector to reject charged particle events.
Details of the LAT instrument and pre-launch expectations of the performance can be found in \cite{Atwood09}. 
Relative to earlier gamma-ray missions, the LAT has a large $\sim$~2.4~sr field of view, a large effective
area ($\sim$~8000~cm$^2$ for $>$1~GeV if on-axis) and improved angular resolution or point-spread function~(PSF, better than 1$^\circ$ for 68\% containment at 1~GeV). 
%Thus, the LAT is ideally suited for detailed spectral studies over four orders of magnitude in energy and source associations of Galactic sources.
%Thus, the LAT is {\bf adequately} suited for detailed spectral studies
%in a wide energy range and source associations of Galactic sources.

%    * Some details about the LAT (from the LAT paper; cite Atwood, W. B. et al. 2009, ApJ arXiv:0902.1089v1):
%          o field of view is ~2.4 sr (at 1GeV)
%          o effective area >1 GeV is ~8000 cm$^2$ on axis
%          o The single event point-spread function (PSF) depends strongly on both the energy and the conversion point in the tracker, but less on the incidence angle. For 1 GeV normal incidence conversions in the upper section of the tracker the PSF is 0.6 degrees
%          o Energy resolution is <15% at energies >100 MeV
%          o LAT observes the entire sky every ~3 h (2 orbits)
%    * The trigger rate is ~2.2 kHz (averaged over many orbits)
%    * The "gamma filter" rate of candidate gamma rays sent to the ground is ~400 Hz (averaged over many orbits)

Routine science operations of the LAT began on August 4, 2008, after the conclusion of a commissioning period.
We have analyzed events in the W28 field, collected from August 4, 2008, to July 4, 2009, with a total exposure of $\sim$~2.8~$\times$~10$^{10}$~cm$^2$~s~(at 1~GeV).
During this time interval, the LAT was operated in sky survey mode nearly all of the time.
In this observing mode the LAT scans the sky, obtaining complete sky coverage every 2 orbits~($\sim$~3~hr) and relatively uniform exposures over time.

We used the standard LAT analysis software, \emph{ScienceTools} version v9r15, which is available from the \emph{Fermi} Science Support Center (FSSC)\footnote{Software and documentation of the \emph{Fermi} \emph{ScienceTools} are distributed by \emph{Fermi} Science Support Center at http://fermi.gsfc.nasa.gov/ssc}, and applied the following event selection criteria: 
a) events have the highest probability of being gamma rays, i.e., they should be classified as so-called Pass 6 \emph{diffuse} class \citep{Atwood09}, 
b) the reconstructed zenith angles of the arrival direction of gamma rays should be less than 105$^\circ$, 
to minimize contamination from Earth albedo gamma rays,
%in order to exclude periods where the Earth enters the LAT field of view, 
c) the center of the LAT field of view should be within 39$^\circ$ from the zenith in order 
to exclude data from the short time intervals when the field of view can be partly occulted by the earth.
% because the Earth limb is frequently in the LAT field of view in this mode,
%and d) all events taken when the spacecraft was within the South Atlantic Anomaly were also excluded.
There are no gamma-ray bursts detected by the LAT within 15$^\circ$ of the W28 field, thus we did not need to apply any additional time cut.
The energy range analyzed here is restricted to $>$~200~MeV to avoid possible large systematic uncertainties due to the strong Galactic diffuse emission near the Galactic center, 
smaller effective area, and much broader PSF at lower energies.

%**** is this true for normal ST and data set ? ***
%The alignment of the LAT pointing direction with
%the celestial frame was calibrated using a large set of high
%latitude gamma-ray sources to better than \citep{Abdo09a}. 
% Systematic uncertainties in the position due to inaccuracies in the
% point-spread function and the telescope alignment are estimated to be < 1¡ë??

\section{ANALYSIS AND RESULTS}
\label{sec:ana}
\subsection{Source position and Spatial Extension}
\label{subsec:extension}
\begin{figure}
%\epsscale{.80}
%\plotone{f1.pdf}
\plotone{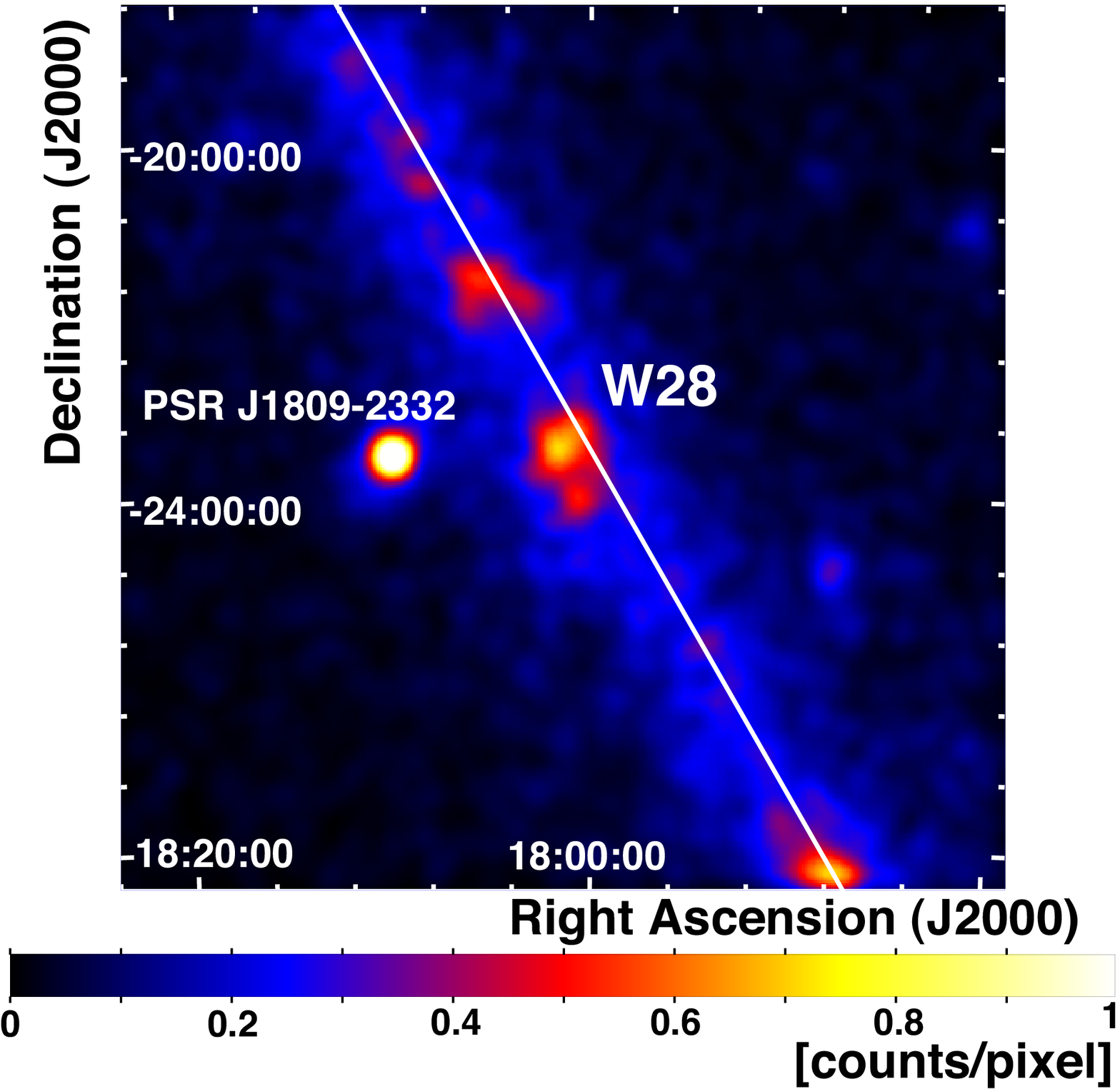}
\caption{\emph{Fermi} LAT 2--10~GeV counts map around the supernova
 remnant~(SNR) W28.
The count map is smoothed by a Gaussian kernel of $\sigma$ $=$
 0.$^\circ$2, with the pixel size of 0.$^\circ$025.
The white line from top left to bottom right indicates the Galactic plane. 
\label{fig:cmap}}
\end{figure}

\begin{figure}
%\epsscale{0.4}
%\plotone{f2a.pdf}
%\epsscale{0.9}
%\plottwo{f2b.pdf}{f2c.pdf}
% \leavevmode
% \includegraphics[scale=.35]{f2a.pdf}
% \includegraphics[scale=.35]{f2b.pdf}
% \includegraphics[scale=.35]{f2c.pdf}
 \includegraphics[scale=.35]{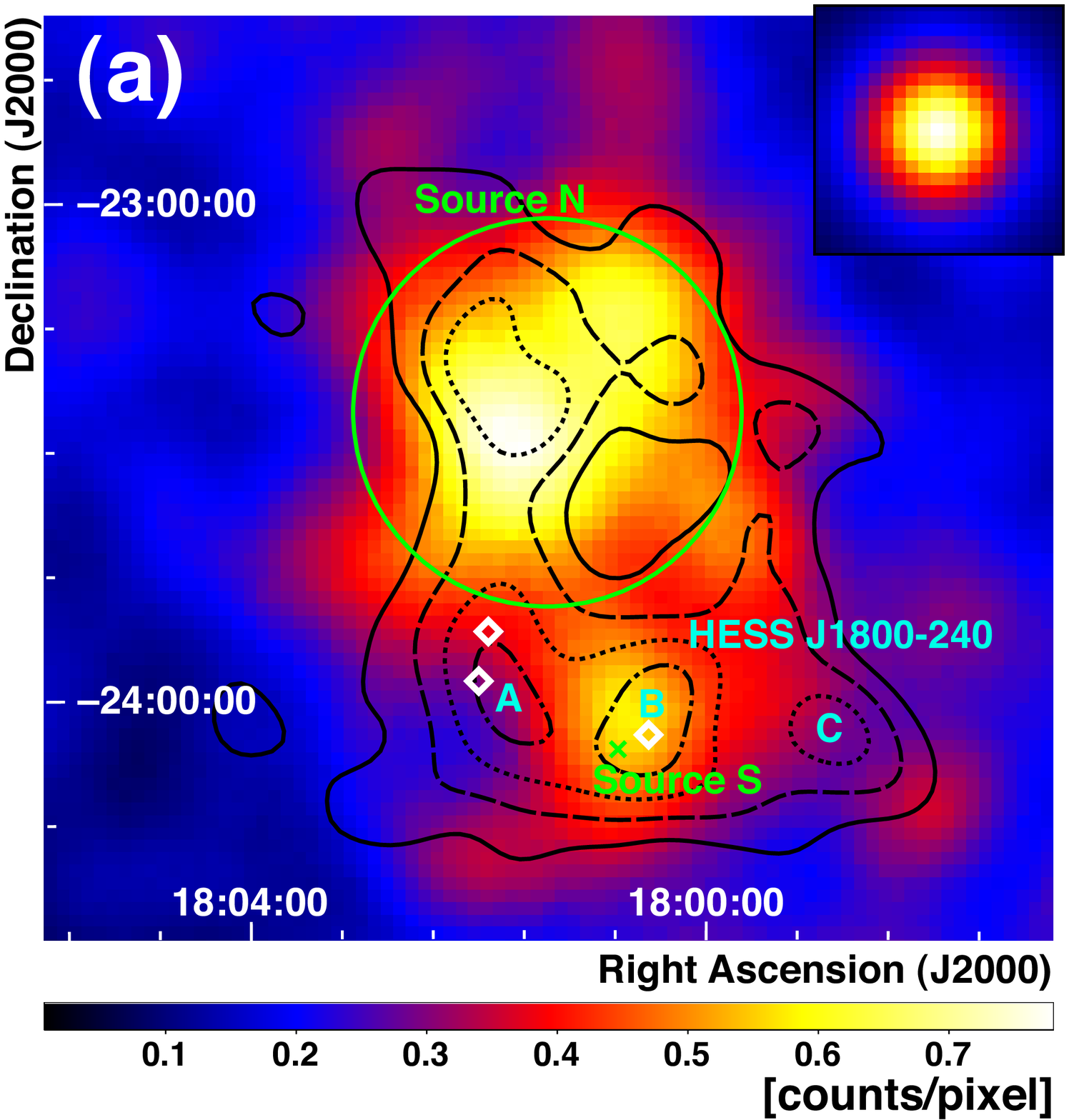}
 \includegraphics[scale=.35]{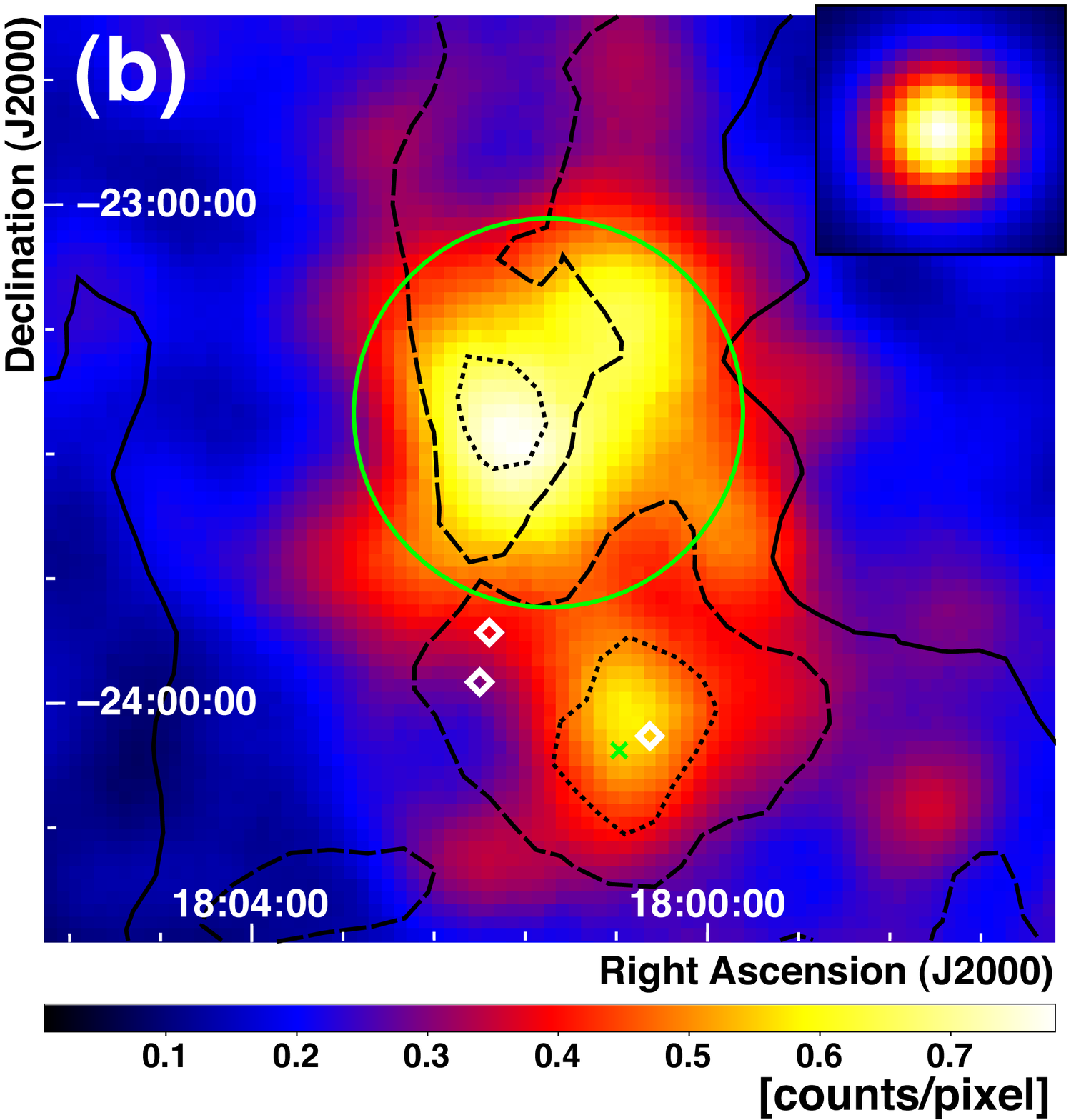}
 \includegraphics[scale=.35]{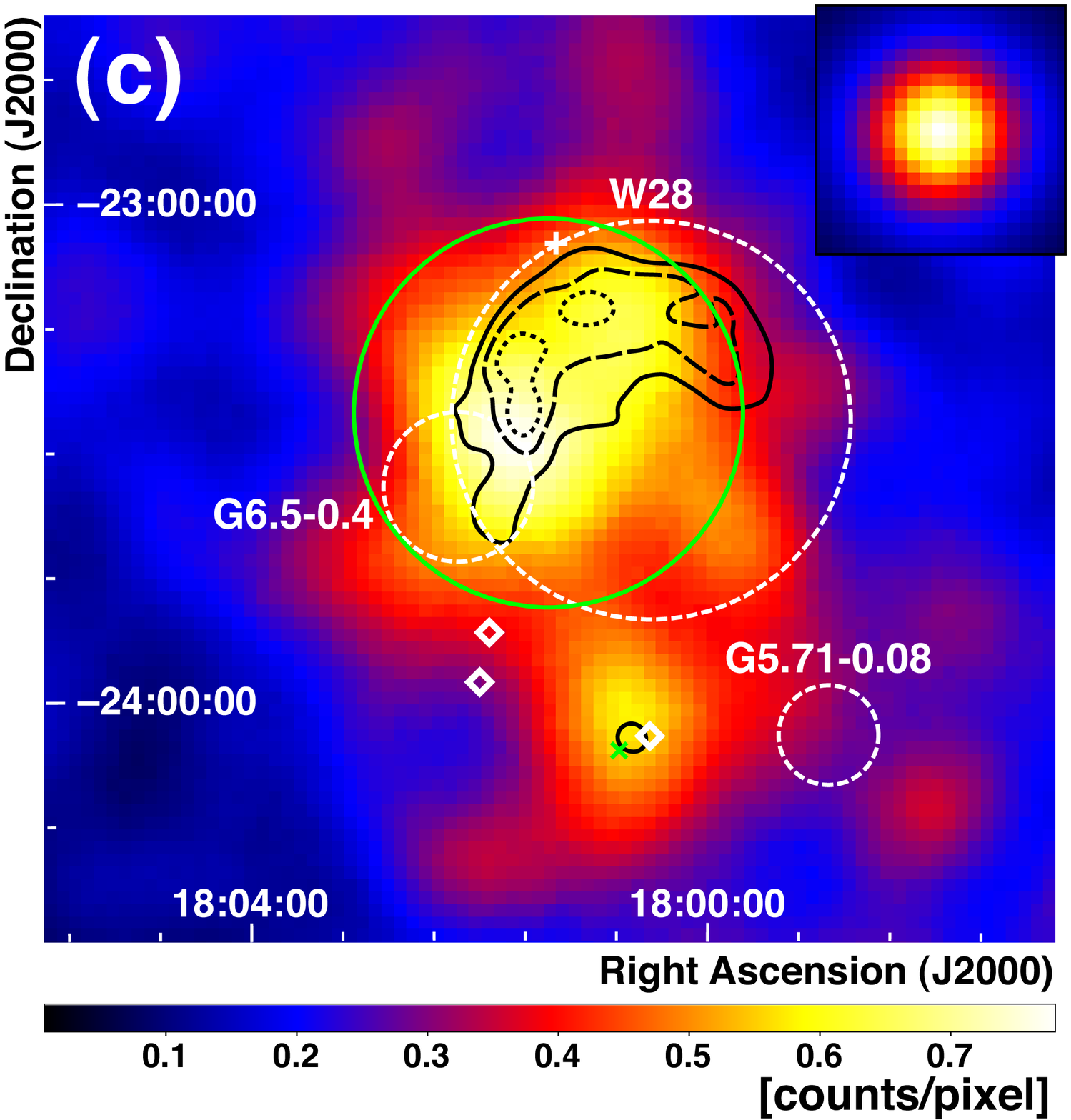}
\caption{Close-up views of the LAT 2--10~GeV count map around W28. 
The counts map is smoothed by a Gaussian kernel of $\sigma$ $=$
 0.$^\circ$2, with the pixel size of 0.$^\circ$025.
The inset of each figure shows 
the effective LAT PSF for a photon spectral index of 2.5.
A green circle in the north of each figure indicates the best-fit disk
 size for \SourceN.
A green cross indicates the position of \SourceS.
 White
diamonds indicate \hii\ regions;
W28A2~(see text), G6.1$-$0.6~\citep{Kuchar97},
 G6.225$-$0.569~\citep{Lockman89}.
The diamond on the right is W28A2.
Black contours in (a) show the H.E.S.S. significance map for TeV gamma rays at 20, 40, 60 and 80\% of the peak value~\citep{Aharonian08}.
Bright TeV spots in the south are HESS J1800-240 A, B and C as indicated in the figure.
Black contours in (b) give CO~(J=1--0) line intensity taken by NANTEN at
 25, 50, 75~\% levels, for the velocity range from 0~km~s$^{-1}$ to
 20~km~s$^{-1}$, corresponding to kinematic distances of approximately 0
 to $\sim$~4~kpc~\citep{NANTEN1,NANTEN2}.
%The large uncertainty of the kinematic distance of the clouds is due to the uncertainty of the Galactic rotation model close to the Galactic center.
Black contours in (c) indicate the VLA 90cm image at 25, 50, 75~\% of the peak intensity~\citep{brogan06}.
Outer boundaries of SNRs, as determined by the radio images, are drawn as white dashed circles. 
A white plus sign shows the position of PSR~J1801$-$23.
\label{fig:cmap2}}
\end{figure}

Figure~\ref{fig:cmap} shows a smoothed count map in the 2--10~GeV energy band in 
a 10$^\circ~\times$~10$^\circ$ region around W28.
Figure~\ref{fig:cmap2} gives comparisons with images of other wavebands in close-up view.
Black contours indicate the H.E.S.S. significance map in (a), CO (J=1-0) line intensity taken by NANTEN for the velocity range from 0~km~s$^{-1}$ to 20~km~s~$^{-1}$ (corresponding to kinematic distances of approximately 0 to 4~kpc) in (b), and a VLA image in (c).
Correlations between GeV gamma rays observed by LAT and some of the H.E.S.S. sources are evident. Gamma rays are also bright in the brightest spots in the CO contours, and bright regions in the gamma-ray image extensively overlap with bright regions in the VLA contours.
%Green contours in Fig.~\ref{fig:cmap}~(b) indicate the significance levels of TeV
%gamma rays at 4, 5 and 6~$\sigma$ observed by H.E.S.S.
%Correlations between GeV gamma rays observed by LAT and some of H.E.S.S. sources are evident.
There are two LAT sources in the vicinity of W28 in the 1FGL catalog \citep{1yrCatalog}: \FGLnorth, and \FGLsouth . 
No obvious pulsations of gamma rays are found in these sources.
Hereafter we refer to \FGLnorth\ 
 as \SourceN, and to \FGLsouth\  as \SourceS. 
%It is worthwhile to note that a possibility of 2 or more gamma-ray sources was suggested by \cite{Torres03} from the existence of multiple molecular clouds and the pulsar PSR B1758-23 around an EGRET source in this region.

In order to quantitatively evaluate the extension and location of these two sources, we apply the maximum likelihood tool, {\tt gtlike}, which is publicly available as part of \emph{Fermi} \emph{ScienceTools}.
The likelihood is the product of the probability of observing the gamma-ray counts of each spatial and energy bin given the emission model, and the best parameter values are estimated by maximizing the likelihood of the data given the model \citep{Mattox96}.
%{\bf maximum likelihood had been used earlier than 1996. I am not sure what's new in Mattox paper}
The probability density function for the likelihood analysis included 
a)~individual sources detected in the preliminary LAT 1-year catalog,
b)~the Galactic diffuse emission resulting from cosmic-ray interactions with the interstellar medium and radiation based on the LAT standard diffuse background model \emph{gll\_iem\_v02} available from FSSC\footnote{The model can be downloaded from 
http://fermi.gsfc.nasa.gov/ssc/data/access/lat/BackgroundModels.html.},
and c) the isotropic component to represent extragalactic and residual
cosmic-ray backgrounds using the isotropic spectral template \emph{isotropic\_iem\_v02} from FSSC.
 Note that we make energy dependent corrections of the the Galactic diffuse model by multiplying a power law function with the spectral index free to vary in the fit.
This correction gives better spectral fits by taking account of local systematic discrepancies between the data and the Galactic diffuse model.
The region of interest for the binned maximum likelihood analysis was a square
region of 20$^\circ$$\times$20$^\circ$ centered on W28 with a pixel size
of 0.$^\circ$1.
The instrument response functions~(IRFs) used in our work were the ``Pass~6~v3'' (P6\_V3) IRFs,
which were developed following launch
 to address gamma-ray detection inefficiencies that are correlated with background rates. 

%The diffuse sources contribute $\sim$****\% of the observed photons shown in
%Fig.~\ref{fig:countmap}.
%The sum of the extra-galactic background, unresolved sources and
%instrumental background: it is assumed to be isotropic in the Galactic coordinate.
%(*** nearby source contribution etc. must be estimated *** contributed
%approximately *** \% to the flux at the location of W28. )
%An alternate fitting method using energy dependent
%regions of interest was used, yielding compatible
%results that were folded into the systematic errors
%*** need ptlike and unfolded ***.
%The Galactic diffuse emission was modeled using GALPROP,
%described in \citep{Strong04}
%and \citep{Strong07}, updated to include recent HI and CO surveys,
%more accurate decomposition into Galactocentric rings,
%and many other improvements, including some from comparison
%with LAT data \citep{Abdo09b}. The GALPROP
%run designation for our model is 54\_59varh7S. 

Using the tools described above, we investigated spatial extensions of the two LAT W28 sources.
Here we used only events above 2.15~GeV to take advantage of narrower PSF
in the higher energy band.
%. and less contamination from the Galactic diffuse component which has rather soft spectrum. 
%=> not correct, as a result of higher PSF,contamination will be reduced
% no, softer galactic diffuse also helps
Before investigating the detailed extension, we first determined the flux and spectral index for all components except for \SourceN\ and \SourceS.
In this process, the normalization of the Galactic diffuse emission and the flux and spectral index of power-law model for the sources within 5 degrees of \SourceN\ were set free to account for the effects of sources around W28 on the background flux in the fit.
% The flux and spectral parameters for source outside the above region
%and position of all sources are set to the parameters from the
%preliminary year-1 LAT catalog. => not 1yr catalog, but 9month, and
%this is described in the previous page
The flux and spectral index except for those of \SourceN\ and \SourceS\ are fixed hereafter.
We used a radially symmetric uniform disk
to evaluate the extension. 
% circular two-dimensional (2D) Gaussian
We varied the radius and location of the disk while holding the position of \SourceS\ 
fixed at the catalog position, 
and evaluated the resulting maximum likelihood value~($L_{\rm ex}$) with respect to the maximum likelihood for no source hypothesis~($L_{\rm 0}$) and the point source hypothesis~($L_{\rm ps}$).
A two-source hypothesis ($L_{\rm 2s}$) is also evaluated by scanning positions of two point sources.

The best likelihood ratio $-2\ln (L_{\rm ps}/L_{\rm ex})\approx 76$ is obtained for \SourceN\ with a disk radius of $\sigma =  0\fdg 39$, which rejects a point source hypothesis at more than 8~$\sigma$.
Note that we validated this procedure by applying this method on a nearby gamma-ray pulsar,
PSR~J1809$-$2332~\citep{LATpulsarCatalog}~($\sim$~2$^\circ$ away from W28),
where we find the extension to be consistent with a point source.
In addition, the best likelihood ratio $-2\ln (L_{\rm 0}/L_{\rm
ex})\approx 336$ for the disk shape is substantially better than that
for two point sources, $-2\ln (L_{\rm 0}/L_{\rm 2s})\approx 276$, 
where the positions of the two sources were free in the optimization.
Therefore we use the best-fit disk model for further analysis of \SourceN.
%This pulsar predominantly emits pulsed gamma rays above 1~GeV, {\it i.e.}
%negligible contamination of an extended source like pulsar wind nebula,
%and has fairly strong emission above 2.15~GeV beyond typical pulsar cutoff energy closely resembling spectral shape of sources in this study.
The best location of the disk model is found at (R.A., Dec.)~$=$~(18${}^h$01${}^m$21${}^s$, $-$23${}^\circ$26$'$26$''$) with an error radius of $0\fdg 03$ at 68~\% confidence level.

The extension of \SourceS\ is also investigated using the same procedure as above. 
We did not find significant extension.
%Therefore, we treat \SourceS\ as a point source hereafter.
An upper limit on the size of the gamma-ray emission was
 obtained by investigating the decrease of the likelihood 
with increasing source size. 
Under the assumption of a disk shape, 
the upper limit amounts to 16$'$ at the 68\% confidence level.
We treat \SourceS\ as a point source hereafter.
The best-fit location of \SourceS\ was estimated to be (R.A., Dec.)~$=$~(18${}^h$00${}^m$59${}^s$, $-$24${}^\circ$11$'$31$''$) using gamma rays above 2.15~GeV.
This location is 13$'$ away from the location of \FGLsouth\ using gamma rays above 0.1~GeV and assuming \FGLnorth\ is a point source.
In order to investigate the origin of the position difference, we performed localization with the same energy cut as the catalog analysis assuming a point source hypothesis and the extended source hypothesis.
We found that the localization with a point source hypothesis is consistent with the catalog position while the extended source hypothesis yields the location consistent with our localization with $E>2.15$~GeV cut.
From this comparison, we conclude that the localization difference is mostly due to the extension of \SourceN. 
%Extensive investigation is required to understand the origin of the offset, which could be due to energy dependence of the spatial distribution, or systematic effect due to uncertainties of the Galactic diffuse model.
%Since further investigation of this matter is beyond the scope of this paper, 
We use the average of two localizations, (R.A., Dec.)~$=$~(18${}^h$00${}^m$47${}^s$, $-$24${}^\circ$05$'$45$''$), for the following analyses, with a systematic uncertainty of half of the positional difference, $\sim$~6$'$.

We find no strong gamma-ray emission in the LAT energy band at the locations of HESS~J1800$-$240~A and C.
We evaluated the likelihood of point sources at these locations and
found likelihood ratios $-2\ln (L_{\rm 0}/L_{\rm ps})\approx 10$
(corresponding to a $\sim$~3~$\sigma$) for
HESS~J1800$-$240A and $-2\ln (L_{\rm 0}/L_{\rm ps})\approx 9$
(corresponding to a $\sim$~3~$\sigma$) for HESS~J1800$-$240C, which confirms no significant emission at these locations.

\subsection{Energy Spectrum}

\begin{figure}
%\epsscale{1.20}
%\plottwo{f3a.pdf}{f3b.pdf}
%\includegraphics[angle=90,scale=.40]{f3a.pdf}
%\includegraphics[angle=90,scale=.40]{f3b.pdf}
\includegraphics[scale=.40]{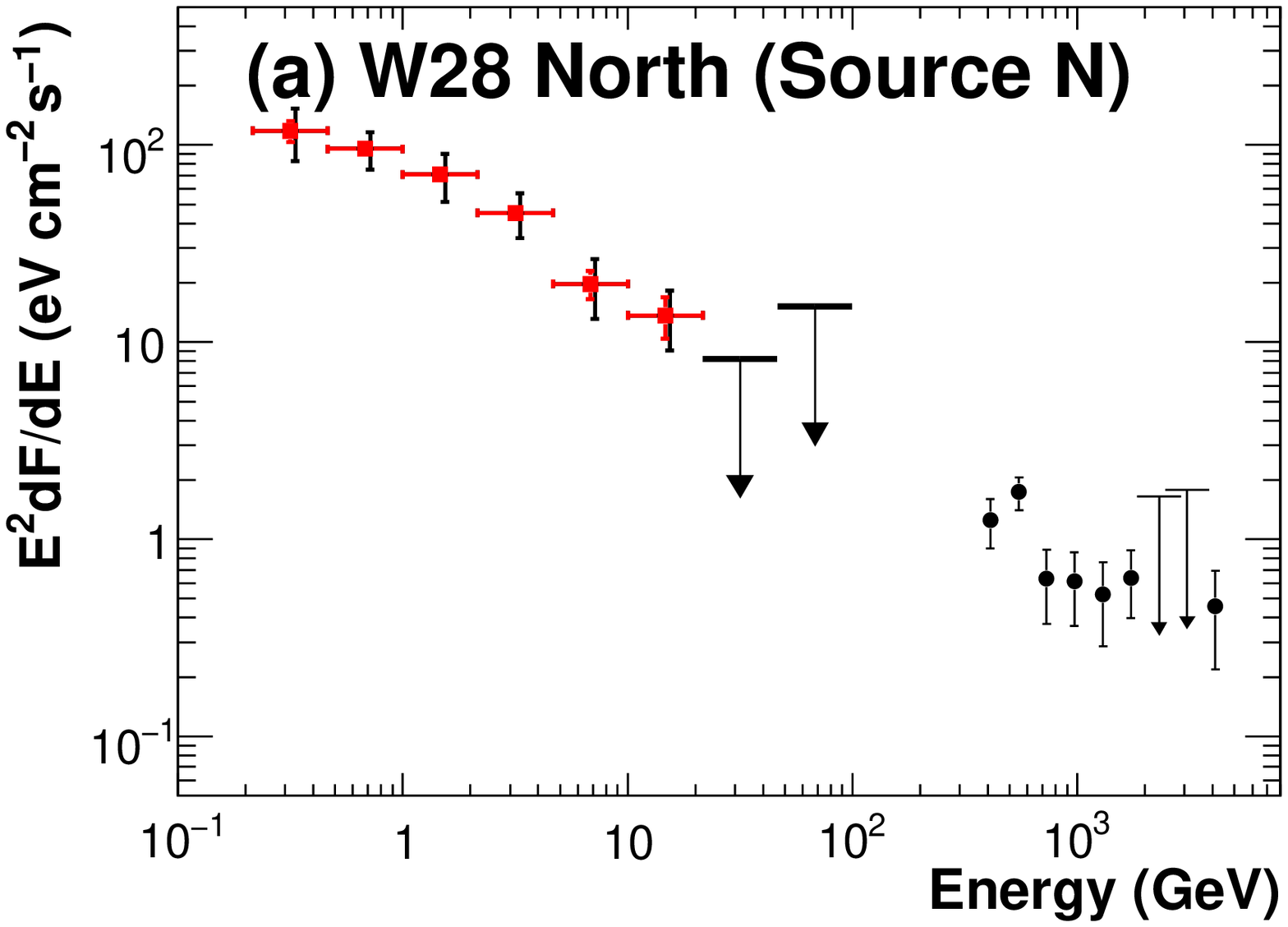}
\includegraphics[scale=.40]{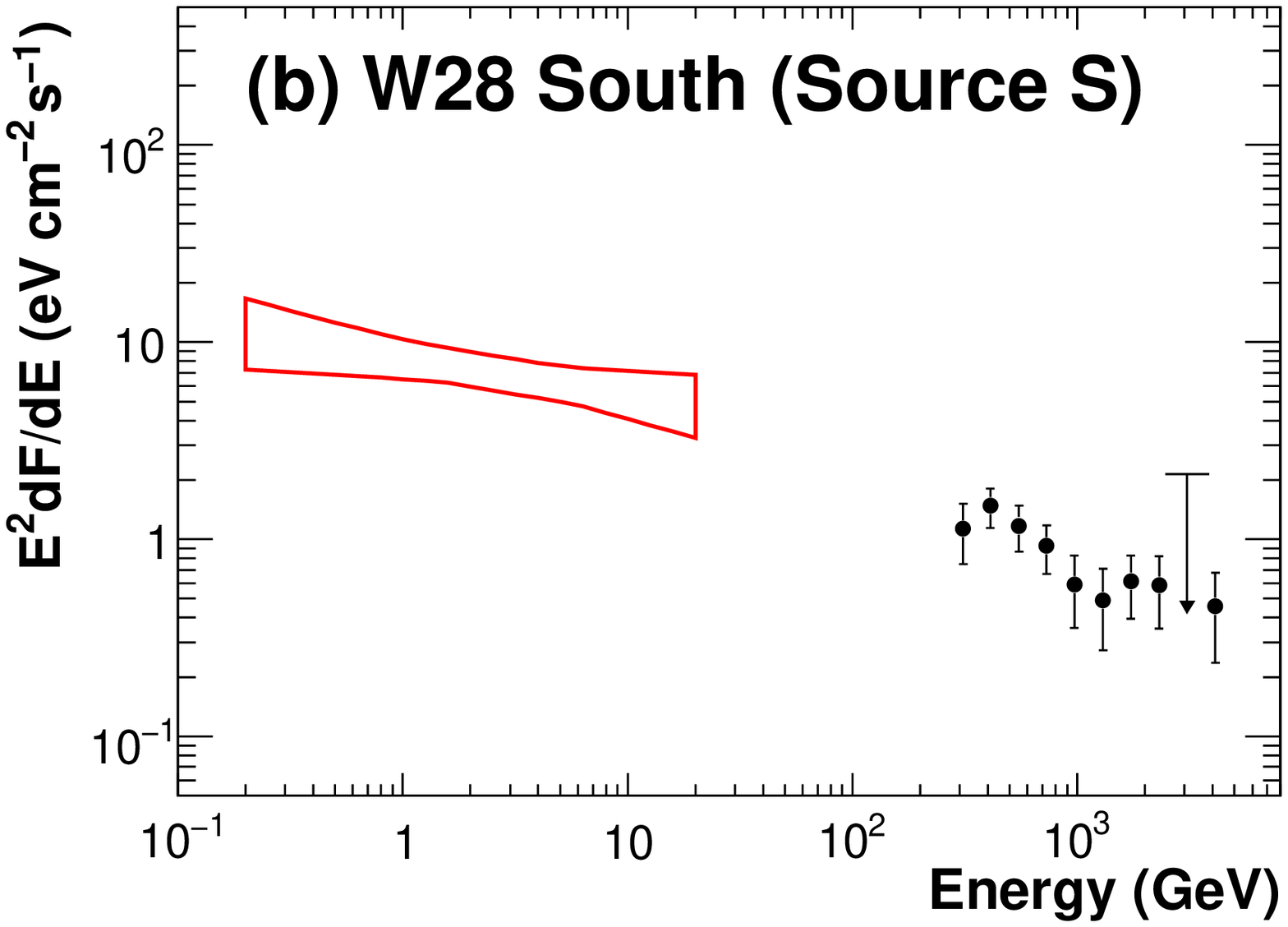}
\caption{ (a)~Spectral energy distribution of the {Fermi} LAT source on
 the northeast boundary of the supernova remnant W28. 
The red squares
in the GeV regime are the LAT data.
Horizontal bars for data points indicate the energy range used in the fit.
Vertical bars show statistical errors in red and systematic errors in
 black. 
Upper limits are obtained at the 90\% confidence level
in energy bins where
the likelihood ratio value is $<$~9.
The black circles represent data points for HESS~J1801$-$233~\citep{Aharonian08}.
 (b)~Spectral energy distribution of the LAT source to the south of W28.
The red region is the 68\% confidence range of the LAT spectrum.
The black circles show data points for HESS~J1800$-$240B~\citep{Aharonian08}.
 \label{fig:spec}}
\end{figure}

We used the maximum likelihood fit tool, {\tt gtlike}, for the spectral analysis of LAT sources in this region.
%In order to produce a spectral energy distribution~(SED) in a model independent manner, fits were performed to obtain the flux in narrow energy bins, which is relatively independent of the spectral index.
Flux points were obtained by performing the maximum likelihood analysis
in each energy bin.
We used eight logarithmically spaced energy bins from 215~MeV to 100~GeV for \SourceN.
%, and three energy bins (200~MeV--1~GeV, 1~GeV--10~GeV, 10~GeV--100~GeV) for \SourceS\ due to poor statistics.
We did not divide the energy range from 0.2~GeV--100~GeV into bins for 
\SourceS\ due to poor statistics.
Fig.~\ref{fig:spec} shows resulting SEDs for (a)~\SourceN\ and (b)~\SourceS.
The 68\% confidence region is illustrated for \SourceS\
 up to $\sim$~20~GeV. We do not have sufficient number of events above $\sim$~20~GeV to
constrain the spectrum.
Upper limits at 90\% confidence level are calculated assuming a photon
index of 2 if the detection is not significant in an energy bin, i.e., the
likelihood ratio with respect to no source is less than 9.
Note that the value of the spectral index has negligible effect on the upper limits.
We take into account systematic errors due to uncertainties of the
extension, the Galactic diffuse model, the LAT effective area,
 and the effect of the nearby gamma-ray pulsar PSR~J1809$-$2332.
Systematic errors associated with the extension are estimated by varying the source size by $\pm~1~\sigma$.
We also evaluated the effect of the shape by comparing the disk shape used in this paper and a circular two-dimensional Gaussian.
Systematic errors due to the Galactic diffuse model are estimated by using the residual gamma-ray data with respect to the best fit model in the region where no LAT source is present, 
specifically,  $l=2\fdg 3-4\fdg 7$ and $b=-1\fdg 25-0\fdg 75$.
The observed residual is energy dependent and can be modeled as $\sim$~(116~($E$/1~GeV)$^{6.45 \times 10^{-2}}-100$)~\% of the total Galactic diffuse flux.
The normalization of the Galactic diffuse model is adjusted according to the above equation to estimate the systematic error on the source flux.
We evaluated the position dependence of this residual and found a dispersion of $\sim$5\%, which is not a large effect.
%We also evaluated systematic error due to effect of bright sources on the normalization of the Galactic diffuse model by shrinking the region of interest to exclude the Galactic center region.
Uncertainties of the Galactic diffuse model make a dominant contribution to 
systematic errors in all the energy bins.
Uncertainties of the LAT effective area is 10\% at 100~MeV, decreasing to 5\% at 500~MeV, and increasing to 20\% at 10~GeV and above~\citep{Rando2009}.
The effect of the nearby gamma-ray pulsar was estimated by varying the flux by $\pm$~1~$\sigma$.

We evaluated a possibility of a spectral break in the LAT energy
band by comparing the likelihood of the spectral fit for the LAT data
between a simple power law and a broken power law as a spectral model of
\SourceN\ and \SourceS.
The fits yield the likelihood ratio  $-2\ln (L_{\rm PL}/L_{\rm
BPL})\approx 62$ for \SourceN,
 where $L_{\rm PL}$ and $L_{\rm BPL}$ are the likelihoods for the simple
 power-law model and the broken power-law model, respectively.
The likelihood ratio slightly decreases to at least $\sim
48$ (corresponding to 6.6~$\sigma$ with two degrees of freedom) 
in the worst case accounting for $1~\sigma$ systematic uncertainties.
Thus we conclude that \SourceN\ has a spectral break at
1.0~$\pm$~0.2~GeV, where the error is dominated by statistics.
Photon indices are 2.09~$\pm$~0.08~(stat)~$\pm$~0.28~(sys) below the break and
2.74~$\pm$~0.06~(stat)~$\pm$~0.09~(sys) above the break.
Note that the photon index above the break is consistent with the H.E.S.S. measurement of $2.66\pm0.27$ \citep{Aharonian08}.
On the other hand, we do not find any evidence of a spectral break for \SourceS.
The photon index is found to be 2.19~$\pm$~0.14~(stat)~$\pm$~0.41~(sys).

\begin{figure}
%\epsscale{1.20}
%\plottwo{f4a.pdf}{f4b.pdf}
%\includegraphics[angle=90,scale=.40]{f4a.pdf}
%\includegraphics[angle=90,scale=.40]{f4b.pdf}
\includegraphics[scale=.40]{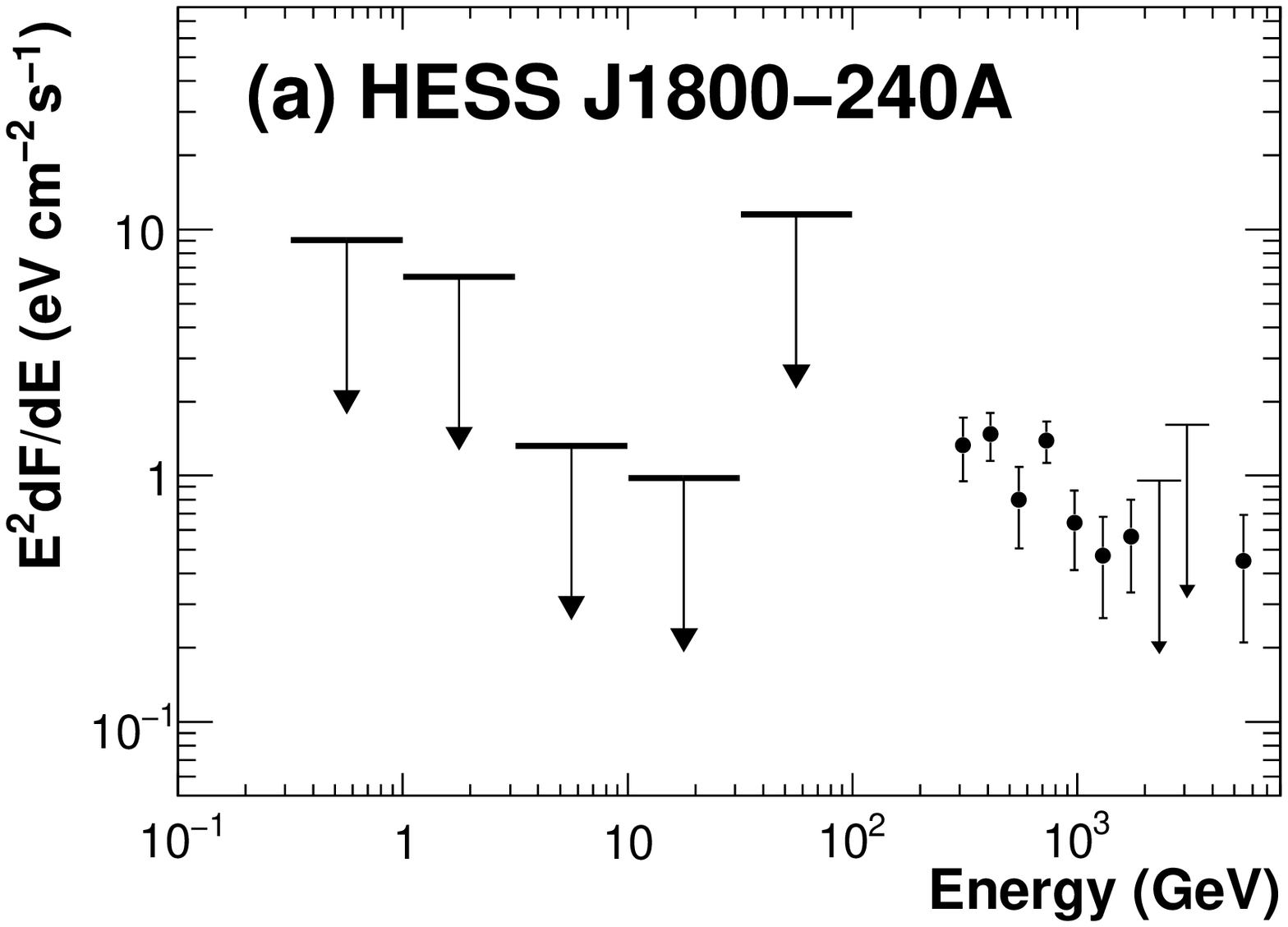}
\includegraphics[scale=.40]{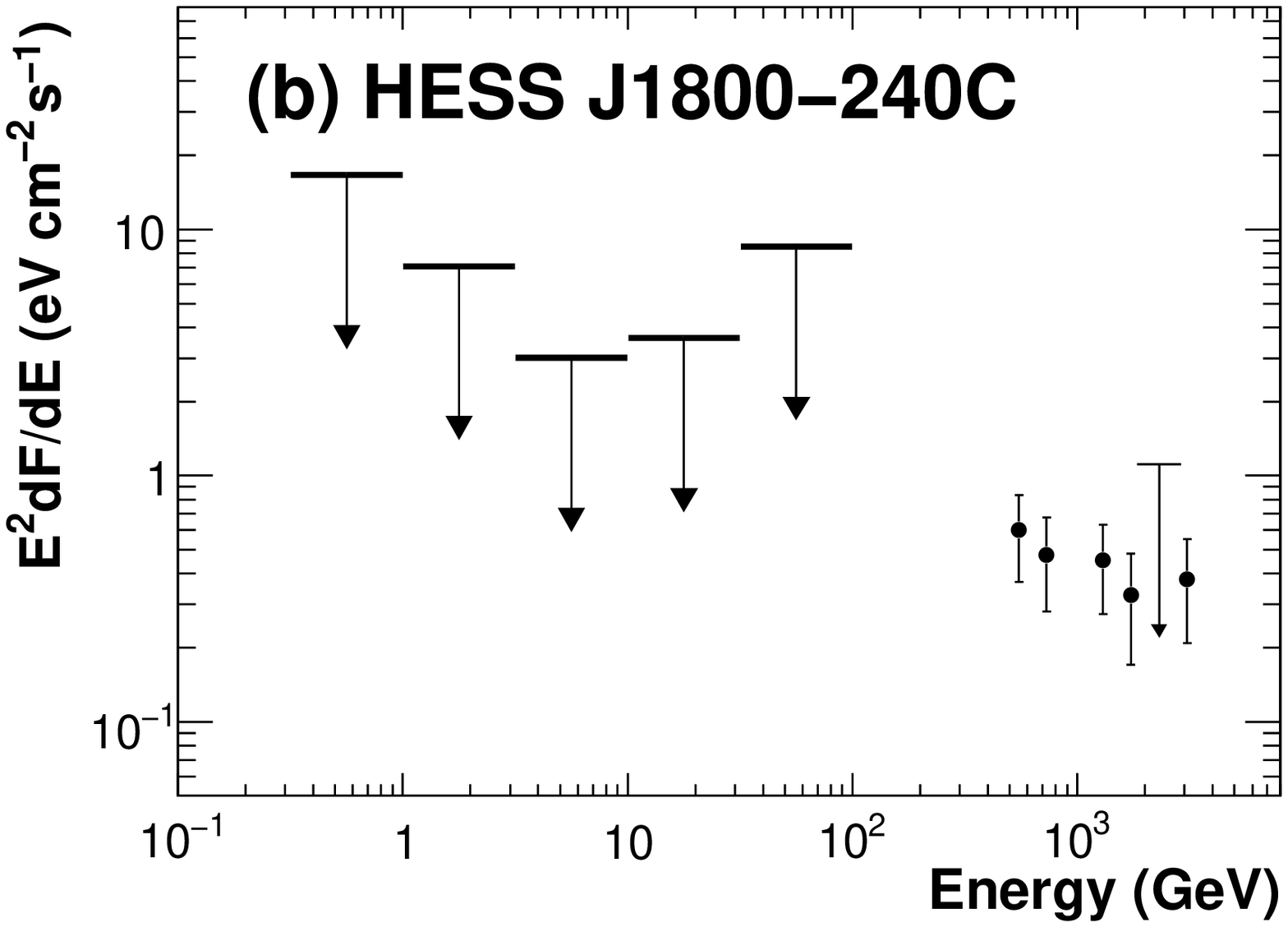}
\caption{Upper limits obtained from the LAT data at the 90\%
 confidence level in each energy bin at the positions of 
(a) HESS~J1800$-$240A and (b) HESS~J1800$-$240C,
 on the assumption of a photon index of 2 for the power-law function.
Black circles are H.E.S.S. spectral measurements~\citep{Aharonian08}.
 \label{fig:spec_hess}}
\end{figure}

We placed upper limits on the gamma-ray flux in the LAT band at the positions of HESS~J1800$-$240A and HESS~J1800$-$240C.
The upper limits are compared with the H.E.S.S. spectra in Fig.~\ref{fig:spec_hess}~(a) and (b).
The gamma-ray upper limits in the GeV band at the location of HESS~J1800$-$240A appear to be inconsistent if the source spectrum is a simple power law.
We evaluated the possibility of a break between the LAT and H.E.S.S. energy band by comparing the likelihood of the spectral fit for the LAT data between a broken power law and a simple power law as a spectral model of this source.
We fixed the spectral index and the flux values using the H.E.S.S. measurements for the simple power-law model while we fixed the spectral index above the break and the flux at the same value as the simple power law and varied the break energy and the spectral index below the break for the broken power-law model.
We used the parameter values at $1~\sigma$ away from the best fit values from H.E.S.S. in the direction where we expect less flux in the LAT energy band, i.e. harder spectral index and lower flux than the best fit values.
%The fits yield the likelihood ratio  $-2\ln (L_{\rm PL}/L_{\rm
%BPL})\approx 33$ (corresponding to a $5.4~\sigma$ significance with two degrees of freedom) for a break energy of $\sim$~100~GeV. 
 The fits yield the likelihood ratio $-2\ln (L_{\rm PL}/L_{\rm
BPL})\approx 35$ (corresponding to a $5.5~\sigma$ significance with two
degrees of freedom) for a break energy between 10~GeV and 100~GeV.
%where $L_{\rm PL}$ and  $L_{\rm BPL}$ are the likelihoods for the broken power-law model and the simple power-law model, respectively.
%The likelihood ratio decreases to $\sim 7$ corresponding to a
%$\sim 2.2~\sigma$ significance if we use parameter values $2~\sigma$
%away from the best-fit values from H.E.S.S.
The likelihood ratio decreases to $\sim 8$ corresponding to a
$\sim 2.4~\sigma$ significance if we use parameter values $2~\sigma$
away from the best-fit values from H.E.S.S.
%We conclude that HESS~J1800$-$240A is likely to have a spectral break
%at around 100~GeV.
We conclude that HESS~J1800$-$240A is likely to have a spectral
break between 10~GeV and 100~GeV.

\section{DISCUSSION}
\label{sec:discuss}
%%% association
% from position
% from extension
%As described above, we have discovered two gamma-ray sources around W28 in the GeV energy band.
\subsection{\FGLnorth\ ~(\SourceN)}
% We have analyzed the two \emph{Fermi} LAT sources in the W28 field, \FGLnorth\ and \FGLsouth .
 
\FGLnorth\ (\SourceN) is found to be extended and positionally coincident with 
HESS~J1801$-$233.
 Given the clear spatial match between the TeV source and the molecular cloud interacting with 
 SNR W28 and a smoothly connected spectrum in the GeV--TeV band, 
we assume that the bulk of GeV and TeV gamma-ray emission comes from the molecular cloud 
illuminated by particles accelerated in supernova shocks. 
%In such cases, ions interacting with ambient matter followed by $\pi^0$-decay 
%or electron bremsstrahlung are the leading candidates that are responsible for the gamma-ray emission.
%Figure~\ref{fig:spec_multi_north}~(a) and (b) show the SED of \SourceN\ with model curves for these two cases.
%The model parameters are summarized in Table~\ref{tab:model}.
Below, we adopt the simplest assumption,
that GeV and TeV gamma rays are emitted by a population of accelerated protons and 
electrons distributed in the same region 
characterized by constant density and magnetic field strength. 
It should be noted that the imaging resolution of the LAT is not good enough to differentiate the GeV emission region from the TeV region. 
We assume the injected electrons have the same momentum distribution as the protons.
This assumption implies a break in the particle momentum spectrum
because the spectral index of the radio data, corresponding to lower
particle momenta, is much harder than that for the gamma-rays, which
correspond to higher particle momenta.
Therefore, we use a broken power-law to describe the particle momentum distribution in the region.
Electrons suffer energy losses due to ionization (or Coulomb scattering), bremsstrahlung, synchrotron processes, and inverse Compton scattering.
The modification of the electron spectral distribution due to such losses was calculated according to \cite{atoyan95}, where electrons are assumed to be injected at $t = 0$ from an impulsive source.
% with the same simple power-law spectrum as protons and transported.
Since diffusive shock acceleration theory generally predicts particle accelerations 
in the Sedov phase with a typical duration $10^3$--$10^4$~yrs, the assumption of an impulsive source would be a good approximation for SNRs with the age of $10^4$--$10^5$~yrs.
We adopt  $4 \times 10^4$~yrs for the age of W28 in this modeling. 
Note that here we consider the minimum momenta of protons and electrons
 to be 100~MeV~$c^{-1}$ since the details of the proton/electron injection process are poorly known.
%Possible older ages of W28 up to $15\times 10^4$~yrs can be accommodated by reducing the magnetic field to avoid too much cooling in TeV energies for electron models.
 The gamma-ray spectrum from $\pi^0$ decay produced by the interaction of protons with ambient hydrogen is scaled by a factor of 1.84 to account for helium and heavy nuclei in target material and cosmic-ray composition~\citep{Mori09}.

\begin{figure}
%\epsscale{1.20}
%\plotone{f5.pdf}
%\includegraphics[angle=90]{f5.pdf}
\includegraphics{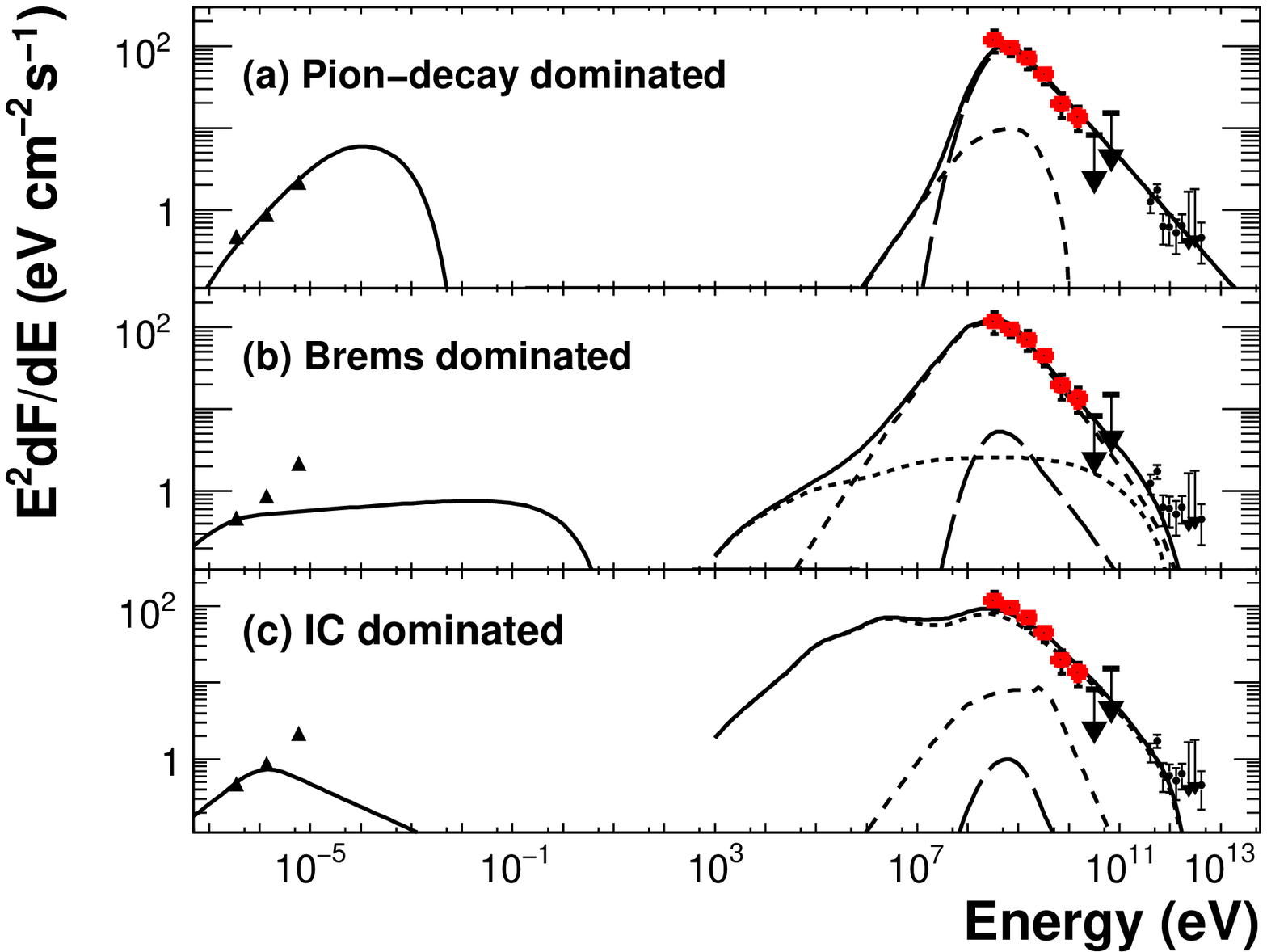}
\caption{Multi-band spectra of the {Fermi} LAT source on
 the northeast boundary of the supernova remnant W28.
 \label{fig:spec_multi_north}
The red squares
in the GeV regime are the LAT data, where
the red and black errors on the flux are statistical and systematic, respectively. 
The radio emission~\citep{Kovalenko94, Dubner00} is modeled by synchrotron
 radiation, 
while the gamma-ray emission is modeled by different combinations of $\pi^0$-decay~(long-dashed curve),
bremsstrahlung~(dashed curve), and inverse Compton~(IC) scattering~(dotted
 curve). 
Details of the models are described in the text.
}
\end{figure}

First, we consider a $\pi^0$-decay model to account for the broadband 
gamma-ray spectrum. Indeed, the hadronic scenario gives the most satisfactory 
explanation for the GeV gamma rays observed in other SNRs interacting 
with molecular gas such as W51C \citep{LAT-W51C} and W44 \citep{LAT-W44}.  
%{\bf A spectral break of \SourceN\ around 1~GeV strongly support this scenario, although it can be formed accidentally by emission of electrons with a break in momentum distribution. }
The number index of protons in the high-energy regime
 is constrained to be $\alpha_{\rm H} \approx 2.7$ from the gamma-ray spectral
 slope. 
 The observed gamma-ray luminosity requires the gas density to be much larger than 
 $\nh\approx 1\ \mathrm{cm}^{-3}$ averaged over the entire SNR shell 
 in order not to exceed the typical kinetic energy of a supernova explosion~($\sim
 10^{51}\ {\rm erg}$).
%{\bf We assume energetic particles are uniformly distributed in the shell-like sturucture with the thickness of 1~pc.
%The radius of the shell is assumed to be 13~pc~($\sim$~0.$^\circ$39 at a distance of 2~kpc), corresponding to the extension of GeV emission.
%The gas density is assumed to be $\nh\approx 10^2\ \mathrm{cm}^{-3}$ based on CO observations.
%The total mass of the gas is $\sim 5.9 \times 10^3$~$M_{\odot}$ considering the angular size of the observation field~($25' \times 25'$).
%This is consistent with the mass of shocked gas given by \cite{Arikawa99}
%Together with the gamma-ray data, we obtain the energy density of cosmic rays to be $\sim$~120~eV~$\mathrm{cm}^{-3}$, which is much higher than that of local cosmic rays.
The resulting total proton energy, $W_{p}\sim 1.3\times10^{49}\cdot(10^2\
\mathrm{cm}^{-3}/\nh)\cdot(d/2\ \mathrm{kpc})^{2}$~erg, is less
than 10\% of the typical kinetic energy of supernova explosions and quite
reasonable.
Note that $W_{p}$ is not the total energy of accelerated protons but
that of the in-situ protons in the molecular clouds.
%{\bf [YU: This is for injected protons or evolved protons? 
%I  prefer to first derive cosmic-ray density and then convert it to the total 
%energy content in the SNR assuming uniform CR distribution.]}
%Relatively high gas density assumed here is quite natural given the correlation of spatial distributions of CO map and gamma rays, and other evidence which supports interaction of the SNR with molecular clouds.
Using the parameters summarized in Table~\ref{tab:model}, 
we calculated radiation model curves as shown in 
 Figure~\ref{fig:spec_multi_north}~(a).
As described above, the spectral index of the proton momentum below the break is determined to be $\alpha_{\rm L}\approx 1.7$ by modeling the radio spectrum as synchrotron radiation by relativistic electrons because we assume that protons and electrons have identical injection spectra.
It is difficult to derive the break point of the proton momentum spectrum from the break of the gamma-ray spectrum since it lies in the region where we expect a gamma-ray spectral curvature due to kinematics of $\pi^0$ production and decays.
Because of this, the gamma-ray spectrum gives an upper bound for the momentum break at $\sim$~5~GeV~$c^{-1}$.
The momentum break cannot be lower than $\sim$~2~GeV~$c^{-1}$ to avoid conflict with the radio data.
Here we adopt 2~GeV~$c^{-1}$.
The magnetic field strength is constrained to be $B \sim 160\;\mu$G, 
for an electron-to-proton ratio of $K_{ep} = 0.01$, which is 
the ratio found in the local cosmic-ray abundance.
Here the ratio is defined at a particle momentum of 1~GeV~$c^{-1}$. 
% since protons are already relativistic and electrons have not been yet cooled radiatively at this energy.
This choice of \kep\ results in minor contribution from the electron bremsstrahlung emission in the gamma-ray band. 
%The radio fluxs at 328~MHz and 1415~MHz are treated as an upper limit
%since it might be contaminated by a thermal
%component~\citep{Dubner00}. % removed by Katagiri (Feb 16 2010)

%The magnetic field strength is constrained to be less than 70~$\mu$G so that the synchrotron radiation does not exceed the radio data.
%Such strong magnetic field is reasonable in molecular clouds.

%with a total energy $W_{\rm e}\approx1.2\times10^{48}\cdot(10^2\ \mathrm{cm}^{-3}/\nh)\cdot(d/2\ \mathrm{kpc})^{2}$~ergs derived from the assumption of \kep.

On the other hand, it is difficult to model the GeV--TeV spectrum by the electron bremsstrahlung component alone because the break in the electron spectrum corresponding to the gamma-ray spectrum will appear in the radio data as shown in Fig.~\ref{fig:spec_multi_north}~(b) although it might be contaminated by a thermal component~\citep{Dubner00}.
%Furthermore, the radio data is not consistent with the electron population distribution described above, 
%can be dominant in the gamma-ray band and represent the data well as shown in Fig.\ref{fig:spec_multi_north}~(b).
%An electron spectral index of $\alpha\approx-2.7$ is obtained from the gamma-ray spectral index.
Moreover, the magnetic field strength is constrained to be less than $B\sim 4\ \mu$G for the age of $4\times 10^4$~yrs since no apparent synchrotron cooling is observed at TeV energies.
The low magnetic field requires the gas density to be smaller than $\nh\sim 5\ \mathrm{cm}^{-3}$ from the flux ratio of the synchrotron component in the radio band and the bremsstrahlung in the gamma-ray band.
These low values for the magnetic field and the gas density are in disagreement with our assumption that the gamma-ray emission comes from the dense molecular cloud.
%since the flux ratio between the
%bremsstrahlung and the synchrotron is proportional to $n/B^2$,
%results in a total electron energy of $W_{\rm e}\approx7.3\times10^{50}\cdot(2\ \mathrm{cm}^{-3}/\nh)\cdot(d/2\ \mathrm{kpc})^{-2}$~ergs to dominate the gamma-ray flux by the bremsstrahlung.
%Please note that the electron-bremsstrahlung dominated model requires $K_{\rm ep}$ far in excess of the local cosmic-ray abundance ratio to suppress the $\pi^0$ decay component, which is very unlikely if the source of particles is the SNR shell.
%Such high $K_{\rm ep}$ is still plausible if the source of particles is a pulsar wind nebula (PWN), however, the total electron energy obtained here is quite large for a PWN.
%In addition, no PWN is observed in this vicinity.
%Therefore, the electron bremsstrahlung is unlikely to be dominant in the gamma-ray band.
%In Fig.~\ref{fig:spec_multi_north}~(b), the SED of \SourceN\ is formally modeled within such a bremsstrahlung scenario.

%The model where the inverse Compton~(IC) of the electrons is the dominant gamma-ray emission, is shown in Fig.\ref{fig:spec_multi_north}~(c).
The gamma-ray spectrum is formally reproduced by inverse Compton (IC) emission from accelerated electrons in Fig.\ref{fig:spec_multi_north}~(c).
The interstellar radiation field for the IC scattering~(see Table~\ref{tab:model}) is comprised of infrared, optical and the cosmic microwave background~(CMB).
The infrared and optical components are taken from the interstellar radiation field at the location of W28 in the GALPROP code~\citep{Porter08}.
Since the spectral shape of the non-CMB component is very complex, it is approximated by two infrared and two optical blackbody components.
The flux ratio between the IC and the synchrotron components 
constrains the magnetic field to be less than $0.6\mu$G.
% We used a broken power-law with a momentum break of
%$\sim$5~GeV~$c^{-1}$ for the electron spectrum to fit the gamma-ray
%spectrum since the IC/synchrotron cooling cannot produce a break in such
%low energies.
%The fairly hard spectral index~($\alpha = 1.5$) below the break is required to suppress the total electron energy below the typical kinetic energy of supernova explosion.
The total energy in electrons is calculated to be $W_{e}\approx9.0\times10^{50}\cdot(d/2\ \mathrm{kpc})^{2}$~erg for an energy density of $\sim$~1.8~eV~cm$^{-3}$ for the interstellar radiation field. 
This model requires a rather low gas density of $\nh \sim 2 \times 10^{-2}~\mathrm{cm}^{-3}$ to suppress the electron bremsstrahlung, which is in contradiction with our assumption.

Observations with \emph{Fermi} have demonstrated that bright, extended gamma-ray sources coincident with middle-aged SNRs interacting with molecular clouds, such as W44 \citep{LAT-W44}, W51C \citep{LAT-W51C}, and IC~443 \citep{LAT-IC443}, exhibit spectral breaks from 9~GeV~$c^{-1}$ to a few tens of GeV~$c^{-1}$ in the proton momentum spectrum.
Our observations of \FGLnorth\ in the vicinity of W28 combined with the radio data constrain the proton momentum break to be in the range, 2--5~GeV~$c^{-1}$. 
The observed energy distribution of relativistic particles (most likely protons) 
could be influenced greatly by diffusive transport of particles. 
If so, the relationship between the observed particle spectrum and the acceleration spectrum 
should be rather complicated. The steep particle spectrum, $\alpha \sim 2.7$, 
deduced for \FGLnorth\ does not necessarily represent the acceleration index. 
\cite{Gabici07} discussed the time evolution of non-thermal emission from molecular clouds illuminated by cosmic rays from a nearby SNR and predicted a steep gamma-ray spectrum for an old SNR due to energy-dependent diffusion of cosmic rays.
More detailed studies of the properties of interactions between SNRs and molecular clouds are required for a comprehensive description of the differences in the break momenta and photon indices above the break among the above SNRs.

%These differences may suggest that the spectral break is not intrinsic to the middle-aged SNRs and depends on the duration of the interaction between the SNR shell and molecular clouds and/or the gas density of the interacting molecular clouds.

% In order to distinguish $\pi^0$-decay component from electron
%bremsstrahlung,
%it is important to reveal below 200~MeV, 
%where the bump feature of gamma rays from $\pi^0$ decays
%due to the interaction of cosmic ray protons with dense matter
%is significant.
%Further study of the systematics are 
%required to investigate the above energy regime,
%which is beyond the scope of this paper.

%The W28 field however is a rich star formation region, and
%several additional/alternative sources of CR acceleration may
%be active. 
%The SNR G7.06$-$0.12 is situated $\sim$~0.$^\circ$7 north of
%HESS~J1801$-$233 and
%on the west side of the \hii\ region M~20. M~20 itself may also
%be an energy source for the molecular clouds in this region. 

\subsection{\FGLsouth\ ~(\SourceS)}
\begin{figure}
%\epsscale{1.20}
%\plotone{f6.pdf}
%\includegraphics[angle=90]{f6.pdf}
\includegraphics{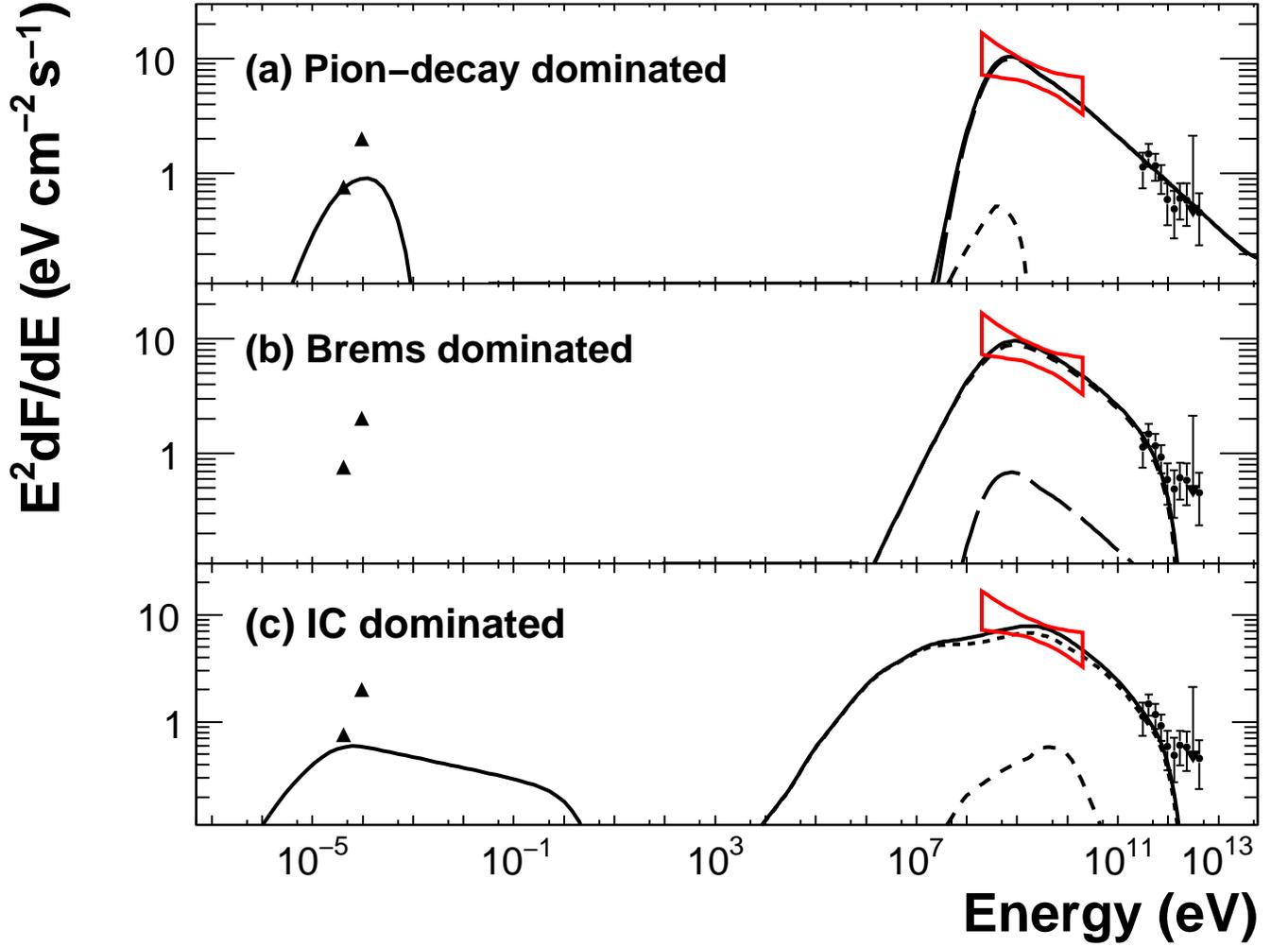}
\caption{Multi-band spectra of the {Fermi} LAT source to the south of the
 supernova remnant W28.
 \label{fig:spec_multi_south}
The red squares
in the GeV regime are the LAT data, where
the red and black errors on the flux are statistical and systematic, respectively. 
The radio emission~\citep{Handa87, gomez91} is modeled by synchrotron
 radiation, 
while the gamma-ray emission is modeled by different combinations of $\pi^0$-decay~(long-dashed curve),
bremsstrahlung~(dashed curve), and inverse Compton~(IC) scattering~(dotted
 curve). 
The sum of the three components is shown as a solid curve.
Details of the models are described in the text.}
\end{figure}

%\FGLsouth\ (\SourceS) is consistent with emission from a point source 
%and spatially coincides with the TeV source HESS~J1800$-$240B, 
%which is associated with 
%molecular clouds that contains the ultra-compact \hii\ region W28A2~(G5.89$-$0.39).
 \FGLsouth\ (\SourceS) was found to have no significant extension and
spatially coincides with the TeV source HESS~J1800$-$240B, 
which is associated with molecular clouds that contain 
 the ultra-compact \hii\ region W28A2~(G5.89$-$0.39).
The observed GeV--TeV spectrum for \SourceS\ can be formally described by $\pi^0$ dominated, bremsstrahlung dominated, and IC dominated models in a similar manners to \SourceN, as shown in Fig.~\ref{fig:spec_multi_south}.
Free-free emission \citep{gomez91} is responsible for 
the radio spectrum of W28A2, which provides only upper limits in the SED.
Thus values of $B$ and $\nh$ are less constrained by  pion and bremsstrahlung models.
We adopt $\nh = 10^3$~cm$^{-3}$ \citep{Aharonian08} for both models.
% Again, leptonic models require high \kep\ value and $\pi^0$ is
%natural explanation of the origin of the gamma rays.
From the positional coincidence, W28A2 can be invoked as a possible 
source of high-energy particles responsible for the gamma-ray emission.
W28A2 exhibits very energetic bipolar molecular outflows \citep{Harvey88, Acord97, Sollins04} which would arise from the accretion of matter by a stellar progenitor.
 \cite{klaassen06} estimated the total kinetic energy of the
outflow of W28A2 to be $3.5\times 10^{46}$~erg, which requires very
high matter density to account for the observed gamma-ray flux in both
pion and bremsstrahlung models.
If the matter density is as high as $\sim$~10$^7$~cm$^{-3}$ as suggested by \cite{klaassen06}, the resulting total energy would be less than a few  percent of the total kinematic energy of the outflow of W28A2.

Cosmic rays that escaped from the SNR W28 in earlier epochs would be another possible explanation given the energetics.
Note that \SourceS\ would lie at a projected distance of $\sim10 \cdot (d/2\, {\rm  kpc})$~pc 
from the southern circular boundary of W28. 
If we assume that cosmic rays are uniformly radiated by the SNR, the cosmic-ray flux bombarding the molecular cloud from the SNR can be scaled by the fraction of the solid angle of the molecular cloud, 
$\sim$~$8 \times 10^{-3}$.
Using this scaling factor, the total energies in the whole SNR W28 could
be calculated from the total energy observed in this source to be
$\sim$~$2 \times 10^{49}$~erg for protons in the $\pi^0$ model.
This is a reasonable value for an SNR.
%{\bf [YU: I prefer to deduce CR density.]} 
The energy density of such cosmic rays is enhanced to be at least
$\sim$~2~eV~cm$^{-3}$ at the molecular clouds 
under the assumption that all of the associated
clouds~($M \sim 4 \times 10^4\; M_{\odot}$ at 2~kpc) are interacting with energetic particles.
The molecular clouds can accumulate those cosmic rays in the past since the diffusion coefficient is expected to be low in dense environments~\citep{Aharonian96}.
% {\bf How about trapping of CRs in MC due to high magnetic field?}

While we detected LAT sources spatially coincident with HESS~J1801$-$233 and HESS~J1800$-$230B, no strong LAT counterpart can be found at the positions of HESS~J1800$-$230A and HESS~J1800$-$230C.
%We found evidence of a spectral break around $\sim$100~GeV for
%HESS~J1800$-$230A.
We found evidence of a spectral break between 10~GeV and 100~GeV for HESS~J1800$-$230A.
The velocity with respect to the local standard of rest of the molecular emission positionally coincident with this source peaks at $\sim$16~km~s$^{-1}$, corresponding to a distance of $\sim$4~kpc \citep{Aharonian08}.
However, the velocity difference between this cloud and the one associated with HESS~J1800$-$230B is only $\sim10$~km~s$^{-1}$ and can be attributed to the proper motion of the cloud complex that contains both clouds.
If the distance to the cloud associated with HESS~J1800$-$230A is 4~kpc, then it is unlikely that this source is associated with W28 since the distance to W28 is $\sim$2~kpc.
\hii\ regions, G6.1$-$0.6 and/or G6.225$-$0.569 could be a source of accelerated particles. 
If the cloud associated with HESS~J1800$-$230A is part of the cloud complex that includes the cloud associated with HESS~J1800$-$230B, then the distance would be 2~kpc, which is very similar to the clouds associated with W28.
In this case, particles escaped from W28 may be responsible for the gamma-ray emissions at both HESS~J1800$-$230A and B and further studies of the origin of the difference in the spectral shapes in these sources may provide good constraints on the particle diffusion process.

\section{CONCLUSIONS}
\label{sec:conclusion}

We have investigated two LAT sources in the W28 field.
 \SourceN\ (\FGLnorth) which is located at the northeast boundary of the SNR W28 is positionally coincident with shocked molecular clouds, and is spatially extended.
The spectrum has a break around 1.0~GeV and smoothly connects to the TeV spectrum, suggesting a physical relationship.
% and a spectral curvature due to decays of $\pi^0$s. }
Decay of $\pi^0$s produced by the interaction of an SNR with molecular clouds naturally explains the gamma rays from \SourceN\ based on the spatial correlation between GeV gamma rays and molecular clouds and the energetics of cosmic rays.
Electron bremsstrahlung can not be ruled out completely although it requires a low density and low magnetic field in contradiction with the association with the molecular clouds.
W28 is the most plausible energy source due to the observational evidence of interaction with the molecular clouds.
The soft spectrum of the gamma rays may be explained by the time evolution of non-thermal emission from molecular clouds illuminated by cosmic rays from a nearby SNR due to energy-dependent diffusion of cosmic rays.

%The \SourceS\ (\FGLsouth) is consistent with the emission from a point source and spatially coincides with the TeV source HESS~J1800$-$240B, molecular clouds, and the ultracompact \hii\ region W28A2~(G5.89$-$0.39).
 
The \SourceS\ (\FGLsouth) was found to have no significant extension 
and spatially coincides with the TeV source HESS~J1800$-$240B, molecular
clouds, and the ultracompact \hii\ region W28A2~(G5.89$-$0.39). 
%Again, leptonic models require high electron/proton ratio and the $\pi^0$ decay is natural explanation of the origin of the gamma rays.
The compact \hii\ region W28A2 and the SNR W28 are possible energy sources, but the W28A2 hypothesis requires extremely dense gas.

%While no significant LAT counterpart is found at the positions of
%HESS~J1800$-$230A and HESS~J1800$-$230C, the LAT upper limits for
%HESS~J1800$-$230A coupled with the H.E.S.S. data points imply a
%spectral break around 100~GeV.
While no significant LAT counterpart is found at the positions of
HESS~J1800$-$230A and HESS~J1800$-$230C, the LAT upper limits for
HESS~J1800$-$230A coupled with the H.E.S.S. data points imply a spectral
break between 10~GeV and 100~GeV.

\acknowledgments

The \textit{Fermi} LAT Collaboration acknowledges generous ongoing support
from a number of agencies and institutes that have supported both the
development and the operation of the LAT as well as scientific data analysis.
These include the National Aeronautics and Space Administration and the
Department of Energy in the United States, the Commissariat \`a l'Energie Atomique
and the Centre National de la Recherche Scientifique / Institut National de Physique
Nucl\'eaire et de Physique des Particules in France, the Agenzia Spaziale Italiana
and the Istituto Nazionale di Fisica Nucleare in Italy, the Ministry of Education,
Culture, Sports, Science and Technology (MEXT), High Energy Accelerator Research
Organization (KEK) and Japan Aerospace Exploration Agency (JAXA) in Japan, and
the K.~A.~Wallenberg Foundation, the Swedish Research Council and the
Swedish National Space Board in Sweden.

Additional support for science analysis during the operations phase is gratefully
acknowledged from the Istituto Nazionale di Astrofisica in Italy and the Centre National d'\'Etudes Spatiales in France.

%JSPA

%\begin{figure}
%\epsscale{.80}
%\plotone{f2.pdf}
%\caption{Radial profiles of the \emph{Fermi} LAT gamma-ray events with the energy of 2.15--4.64~GeV, after the diffuse component is subtracted.
%The reference point of the radial profile is (R.A., Dec.)~=~~(18${}^h$01${}^m$23${}^s$, $-$23${}^\circ$25$'$08$''$), which is the centroid of the best-fit symmetric two-dimensional~(2D) Gaussian spatial model ($\sigma~=~$0.$^\circ$23) on the northeast boundary of the supernova remnant W28.
%Black circles with error bars show the data. 
%A solid line represents the sum of the models, including the best-fit 2D Gaussian source and the point source to the south of W28. 
%Details of the models are described in the text. 
%Dotted lines show individual source components. 
%A expected profile for the point-source model is shown by a hatched histogram for comparison, where the spectral index is assumed to be $-$2.5. \label{fig:radial_profile}}
%\end{figure}

\begin{table}
\begin{center}
\caption{Model parameters for the {Fermi} LAT sources around W28.\label{tab:model}}
\begin{tabular}{lccccccccc}
\tableline\tableline
%\multicolumn{3}{l}{test}
%\multicolumn{3}{l}{test}
 Model  & \kep\tablenotemark{a} &  $\alpha_{\rm L}$\tablenotemark{b} & $p_{\rm b}$\tablenotemark{c} & $\alpha_{\rm H}$\tablenotemark{d} & $B$ &
 $\nh$\tablenotemark{e} & $W_{p}$\tablenotemark{f} & $W_{e}$\tablenotemark{f} \\
   &    &  & (GeV~$c^{-1}$) & &  ($\mu$G) & (cm$^{-3}$) & ($10^{49}$~erg) & ($10^{49}$~erg) \\
\tableline
\SourceN & & & & & & & & & \\
  (a)~Pion & 0.01  & 1.7 & 2 & 2.7 & 160 & 100 & 1.3 & 1.9~$\times$~$10^{-2}$ \\
(b)~Bremsstrahlung & 1 & 1.7 & 1 & 2.7 & 4 & 5  & 1.9 & 4.9  \\
(c)~Inverse Compton\tablenotemark{g} & 1 & 1.7 & 5 & 3.6 & 0.6 & 0.02 & 54 & 90 \\
\SourceS & & & & & & & & & \\
 (a)~Pion & 0.01 & 1.7 & 2 & 2.4 &  1.2~$\times$~$10^3$ & $10^3$ & 1.5~$\times$~$10^{-2}$ & 6.5~$\times$~$10^{-5}$ \\
(b)~Bremsstrahlung & 1 & 1.7 & 1 & 2.4 & 4 & $10^3 $ & 1.1~$\times$~$10^{-3}$  &  2.1~$\times$~$10^{-3}$ \\
(c)~Inverse Compton\tablenotemark{g} & 1 & 1.7 & 15 & 3.2  & 2 & 0.05 &
 1.7 &  2.3 \\
\tableline
\end{tabular}
\tablenotetext{a}{The ratio of electron and proton distribution functions
at 1~GeV~$c^{-1}$.}
\tablenotetext{b}{
A momentum distribution of particles
 is assumed to be a broken power-law, where the indices and the break
 momentum are identical for both accelerated protons and electrons.
The $\alpha_{\rm L}$ is a photon index below the momentum break.}
\tablenotetext{c}{The $p_{\rm b}$ is a momentum break for particle distribution.}
\tablenotetext{d}{The photon index for broken power-law functions
 above the momentum break.}
%\tablenotetext{e}{The exponential cutoff momentum of particle distributions.}
\tablenotetext{e}{Average hydrogen number density of ambient medium.}
\tablenotetext{f}{The distance is assumed to be 2~kpc. The total energy
 is calculated for particles $>$~100~MeV~$c^{-1}$.}
\tablenotetext{g}{Seed photons for inverse Compton scattering of
 electrons include the CMB,
 two infrared~($T_{\rm IR} = 29, 4.9 \times 10^2$~K, $U_{\rm IR} = 0.29,
 5.3 \times 10^{-2}$~eV~cm$^{-3}$, respectively), and
 two optical components~($T_{\rm opt} = 3.6 \times 10^3, 1.0 \times
 10^4$~K, $U_{\rm opt} = 0.37, 0.13$~eV~cm$^{-3}$, respectively) in the
 vicinity of W28, assuming a distance of 2~kpc.}
\end{center}
\end{table}


\begin{thebibliography}{}

%\bibitem[Abdo et al.(2009a)]{Abdo09}
%Abdo, A. A., et al. (The \emph{Fermi} LAT Collaboration) 2009a, \apjs, 183, 46


\bibitem[Abdo et al.(2009)]{LAT-W51C}
Abdo, A. A., et al. (The \emph{Fermi} LAT Collaboration) 2009, \apjl, 706, L1


\bibitem[Abdo et al.(2010a)]{LAT-W44}
Abdo, A. A., et al. (The \emph{Fermi} LAT Collaboration) 2010a, Science in press 

\bibitem[Abdo et al.(2010b)]{LAT-IC443}
Abdo, A. A., et al. (The \emph{Fermi} LAT Collaboration) 2010b, to
			    appear in \apj, arXiv:1002.2198 

\bibitem[Abdo et al.(2010c)]{LATpulsarCatalog}
Abdo, A. A., et al. (The \emph{Fermi} LAT Collaboration) 2010c, to appear in \apjs, arXiv:0910.1608

\bibitem[Abdo et al.(2010d)]{1yrCatalog} 
Abdo, A. A., et al. (The \emph{Fermi} LAT Collaboration) 2010d,
			    submitted to \apss, arXiv:1002.2280 

\bibitem[Acciari et al.(2009)]{Acciari09}
Acciari, V. A., et al. 2009, \apjl, 698, L133

\bibitem[Acord et al.(1997)]{Acord97}
Acord, J. M., Walmsley, C. M., \& Churchwell, E.
1997, \apj, 475, 693


\bibitem[Aharonian \& Atoyan(1996)]{Aharonian96}	
Aharonian, F. A., Atoyan, A. M.
1996, \aap, 309, 917

%\bibitem[Aharonian \& Atoyan(1999)]{Aharonian99}	
%Aharonian, F. A.; Atoyan, A. M.
%1999, \aap, 351, 330

\bibitem[Aharonian et al.(1994)]{Aharonian94} Aharonian, F.~A., Drury, L.~O., \&
			   V$\ddot{\rm o}$lk, H.~J.\ 1994, \aap, 285, 645 

%\bibitem[Aharonian et al.(2004)]{Aharonian04}
% Aharonian, F., et al. (The H.E.S.S. Collaboration) 2004, Nature, 432, 75

%\bibitem[Aharonian et al.(2005)]{Aharonian05a}
% Aharonian, F., et al. (The H.E.S.S. Collaboration) 2005, \aap, 437, L7

%\bibitem[Aharonian et al.(2005)]{Aharonian05b}
% Aharonian, F., et al. (The H.E.S.S. Collaboration) 2005, Science, 307, 1938

%\bibitem[Aharonian et al.(2006)]{Aharonian06}
% Aharonian, F., et al. (The H.E.S.S. Collaboration) 2006, \aap, 449, 223

%\bibitem[Aharonian et al.(2007a)]{Aharonian07a}
% Aharonian, F., et al. (The H.E.S.S. Collaboration) 2007a, \aap, 464, 235

%\bibitem[Aharonian et al.(2007b)]{Aharonian07b}
% Aharonian, F., et al. 2007b, ApJ, 661, 236

\bibitem[Aharonian et al.(2008)]{Aharonian08}
			    Aharonian, F., et al.
			    2008, \aap, 481, 401

%\bibitem[Albert et al.(2006)]{Albert06}
%Albert J., et al. (The MAGIC Collaboration) 	
%	2006, \apjl, 643, L53A

\bibitem[Albert et al.(2007)]{Albert07}
Albert J., et al. (The MAGIC Collaboration) 
			    2007, \apjl, 664, L87

\bibitem[Arikawa et al.(1999)]{Arikawa99}
Arikawa, Y., Tatematsu, K., Sekimoto, Y., \& Takahashi, T.
1999, \pasj, 51, L7

\bibitem[Atoyan et al.(1995)]{atoyan95}
Atoyan A. M., Aharonian A. F., \& V$\ddot{\rm o}$lk, H. J. 1995,
Phys. Rev. D, 52, 3265

\bibitem[Atwood et al.(2009)]{Atwood09}
Atwood, W. B., et al. (The \emph{Fermi} LAT Collaboration) 2009, 
\apj, 697, 1071

%\bibitem[Berezhko \& V\"olk(2006)]{Berezhko06}
% Berezhko, E. G. \& V$\ddot{\rm o}$lk, H. J. 2006, \aap, 451, 981

\bibitem[Blandford \& Eichler(1987)]{blandford87}
 Blandford, R. D. \& Eichler, D. 1987, Phys. Rep., 154, 1

\bibitem[Brogan (2006)]{brogan06}
Brogan C.~L., Gelfand, J.~D., Gaensler, B.~M., Kassim, N.~E., \& Lazio, T.~J.~W.\ 2006, \apjl, 639, L25

\bibitem[Buckley et al.(1998)]{Whipple98} Buckley, J.~H., et al.\ 1998, \aap, 329, 639 

%\bibitem[Chevalier(1999)]{Chevalier99}
%Chevalier R. A., 1999, \apj, 511, 798

\bibitem[Claussen et al.(1997)]{Claussen97}
Claussen, M. J., Frail, D. A., Goss, W. M., \& Gaume, R. A.
1997, \apj, 489, 143

\bibitem[Claussen et al.(1999)]{Claussen99}
Claussen, M. J., Goss, W. M., Frail, D. A., \& Desai, K.
1999, \apj, 522, 349

\bibitem[Dubner et al.(2000)]{Dubner00}
Dubner, G. M., Vel$\acute{a}$zquez, P. F., \& Goss, W. M.,
2000, \apj, 120, 1933

%\bibitem[Esposito et al.(1996)]{esposito96}
%Esposito, J. A., Hunter, S. D., Kanbach, G., \& Sreekumar, P. 1996, ApJ,
%			    461, 820

\bibitem[Frail et al.(1994)]{Frail94}
Frail, D. A., Goss, W. M., \& Slysh, V. I.
1994, \apjl, 424, L111

%\bibitem[Gabici et al.(2009)]{Gabici09}
%Gabici, S., Aharonian, F. A., \& Casanova, S.
%2009, \mnras, 396, 1629

%\bibitem[Gabici et al.(2007)]{Gabici07}
%Gabici, S., Aharonian, F. A., \& Blasi, P.
%2007, Astrophys. and Space Sci., 309, 365

\bibitem[Gabici 
\& Aharonian(2007)]{Gabici07} Gabici, S., \& Aharonian,
			   F.~A.\ 2007, \apjl, 665, L131 
%\bibitem[Ginzburg \& Syrovatskii(1964)]{ginzburg64}
% Ginzburg, V. L. \& Syrovatskii, S. I. 1964, The Origin of Cosmic Rays
% (New York: Macmillan)

\bibitem[G$\acute{\rm o}$mez et al.(1991)]{gomez91}
G$\acute{\rm o}$mez, Y. , Rodr$\acute{\rm i}$guez, L. F.,  Garay, G., \& Moran, J. M.
	1991, \apj, 377, 519

\bibitem[Goudis(1976)]{goudis76}
 Goudis, C. 1976, \apss, 40, 91

\bibitem[Handa et al.(1987)]{Handa87} Handa, T., Sofue, Y., 
Nakai, N., Hirabayashi, H., \& Inoue, M.\ 1987, \pasj, 39, 709 

\bibitem[Hartman et al.(1999)]{Hartman99}
Hartman, R. C., et al. 1999, \apjs, 123, 79

\bibitem[Harvey \& Forceille(1988)]{Harvey88}	
Harvey, P. M., \& Forveille, T.
1988, \aap, 197, L19

%% \bibitem{Long (1991)}
\bibitem[Lozinskaya(1974)]{Lozinskaya74} Lozinskaya, T.~A.\ 1974, 
Soviet Astronomy, 17, 603

%\bibitem[Lozinskaya(1981)]{lozinskaya81}
% Lozinskaya, T. A. 1981, Sov. Astron. Lett. 7, 17

\bibitem[Lozinskaya(1992)]{Lozinskaya92}
 Lozinskaya, T. A., Supernovae and Stellar Wind in the Interstellar
			    Medium (New York: AIP)

\bibitem[Mattox et al.(1996)]{Mattox96}
Mattox, J. R., et al. , 1996, \apj, 461, 396


\bibitem[Mizuno
\& Fukui(2004)]{NANTEN1} Mizuno, A., \& Fukui, Y.\ 2004,
                           Milky Way Surveys: The Structure and
                           Evolution of our Galaxy, 317, 59

\bibitem[Mori(2009)]{Mori09}
Mori, M., 2009, Astropart. Phys., 31, 341

\bibitem[Kaspi et al.(1993)]{Kaspi93}
Kaspi, V. M., Lyne, A. G., Manchester, R. N., Johnston, S., D'Amico, N.,
			    \& Shemar, S. L.	
1993, \apj, 409, L57

\bibitem[Klaassen et al.(2006)]{klaassen06}
Klaassen, P. D., Plume, R.,  Ouyed, R., von Benda-Beckmann, A. M., \&
			    Di Francesco, J. 2006, \apjl, 648, L1079

\bibitem[Kovalenko et al.(1994)]{Kovalenko94} Kovalenko, A.~V., 
Pynzar', A.~V., \& Udal'Tsov, V.~A.\ 1994, Astronomy Reports, 38, 95 


\bibitem[Kuchar
\& Clark(1997)]{Kuchar97} Kuchar, T.~A., \& Clark, F.~O.\ 1997, \apj, 488, 224


\bibitem[Lockman(1989)]{Lockman89} Lockman, F.~J.\ 1989, \apjs,
71, 469


\bibitem[Pittori et al.(2009)]{AGILEcatalog} Pittori, C., et al.\ 2009, \aap, 506, 1563

\bibitem[Porter et al.(2008)]{Porter08}
Porter, T., et al. 2008, \apj, 682, 400

\bibitem[Rando et al. (2009)]{Rando2009} Rando, R. et al., arXiv:0907.0626


\bibitem[Reach et al.(2005)]{Reach05}	
Reach, W. T., Rho, J., \& Jarrett, T. H.
2005, \apj, 618, 297

\bibitem[Rho \& Borkowski(2002)]{Rho02}
Rho, J., \& Borkowski, K. J. 
2002, \apj, 575, 201

%\bibitem[Torres et al.(2003)]{Torres03}
%Torres, D. F., Romero, G. E., Dame, T. M., Combi, J. A., \& Butt, Y. M.
%\physrep, 382, 303.

\bibitem[Sollins et al.(2004)]{Sollins04}
Sollins, P. K., et al. 
2004, \apj, 616, L35

%\bibitem[Sturner et al.(1997)]{sturner97}
%Sturner, S. J., Skibo, J. G., Dermer, C. D., \& Mattox, J. R. 1997, \apj, 490, 619


\bibitem[Takeuchi et al.(2010)]{NANTEN2}
Takeuchi, T. et al.
2010, \pasj, in press 

\bibitem[Vel$\acute{\rm a}$zquez et al. (2002)]{Velazquez02}
Vel$\acute{\rm a}$zquez, P. F., Dubner, G. M., Goss, W. M., \& Green, A. J.
2002, \aj, 124, 2145

\bibitem[Wootten(1981)]{Wootten81}
Wootten, A. 1981, \apj, 245, 105

\end{thebibliography}
\end{document}